\documentstyle[aps,prd,preprint]{revtex}
\begin{document}

\draft

\title{Flavour and Spin of the Proton \\
and the Meson Cloud}

\author{H. Holtmann ${}^1$, A. Szczurek ${}^2$ and J. Speth ${}^1$}
\address{
${}^1$ Institut f\"ur Kernphysik, Forschungszentrum J\"ulich, \\
53425 J\"ulich, Germany \\ ${}^2$ Institute of Nuclear Physics,\\
PL-31-342 Krakow, Poland}

\date{\today}
\maketitle

\begin{abstract}
We present a complete set of formulas for longitudinal momentum
distribution functions (splitting functions) of mesons in the nucleon.
It can be applied in the framework of convolution formalism to the
deep-inelastic structure functions (quark distributions) of the
nucleon viewed as a system composed of virtual 'mesons' and 'baryons'.
Pseudoscalar and vector mesons as well as octet and decuplet baryons
are included.  In contrast to many approaches in the literature the
present approach ensures charge and momentum conservation by the
construction.  We present not only spin averaged splitting functions
but also helicity dependent ones, which can be used to study the spin
content of the nucleon. The cut-off parameters of the underlying form
factors for different vertices are determined from high-energy
particle production data. We find an universal cut-off parameter for
processes involving octet baryons. This information allows one to
calculate the flavour and spin content of the nucleon.  The value of
the Gottfried Sum Rule obtained from our model ($S_G = 0.224$) nicely
agrees with that obtained by the NMC.  In addition, we calculate the
$x$-dependence of the $\overline{d} - \overline{u}$ asymmetry and get
an impressive agreement with a recent fit of Martin-Stirling-Roberts.
The calculated axial coupling constants for semileptonic decays of the
octet baryons agree with the experimental data already with $SU(6)$
wave function for the bare nucleon. As a consequence the Bjorken Sum
Rule is nicely reproduced. Although we get improvements for the
Ellis-Jaffe Sum Rules for the proton and neutron in comparison to the
naive quark model, the MCM is not sufficient to reproduce the
experimental data.
\end{abstract}


\section{Introduction}

It is taken for granted that the nucleon consists of quarks and
gluons and the underlying theory describing their interaction is
QCD.  However, solving the QCD equations for the many-body system at
large distances is at the present time still beyond our practical
abilities.
Deep-inelastic scattering (DIS) of leptons is known to be a very
good tool to study the structure of the proton. Here in the Bjorken
limit one directly tests the quark distributions in the
nucleon.
Although the perturbative regime seems to be well under
control and the QCD evolution equations turned out to be very
successful in relating the quark distributions at different momentum
scales, the nonperturbative effects are poorly known. Recent
experiments on deep-inelastic scattering of leptons from nucleons
\cite{A91,A89} have shown the incompleteness of our understanding of
the proton structure.

The violation of the Gottfried Sum Rule observed by NMC \cite{A91}
\begin{equation}
S_G=\int_0^1 [F_2^p(x)-F_2^n(x)]{dx\over x}
   = 0.24 \pm 0.016
\end{equation}
indicates that the nucleon sea is not flavour symmetric.  Recent fits
of quark distributions to the world data for deep-inelastic and
Drell-Yan processes confirm the asymmetry \cite{MSR93,MSR94}.
Also the new results of the CERN dedicated NA51 experiment \cite{B94}
on the dilepton production in proton-proton and proton-deuteron
scattering gives evidence for the asymmetry.
More detailed information can be expected from the experiments
planned at Fermilab \cite{G92,SSG94}.
Perturbative effects
\cite{RS79} cannot explain the observed asymmetry.
The dressing of the nucleon with virtual mesons provides a natural
explanation
for the excess of $\overline{d}$ over $\overline{u}$ quarks in the
proton \cite{HS91,SS93,Z92,K291} and explains the result of the NA51
experiment \cite{HNSS94}.

Another intriguing result, involving the spin structure of the
proton, was found by the EMC collaboration. Their measured value of
the Ellis-Jaffe Sum Rule \cite{A89} caused great excitement. After a
smooth Regge extrapolation of the data they have found
\begin{equation}
S_{EJ}^{p}
= \int_0^1 g^p_1(x)dx=0.126 \pm 0.010 (stat) \pm 0.015(syst).
\label{EMC}
\end{equation}
This value is more then two standard deviations away from the
original Ellis-Jaffe prediction ($S_{EJ}^p = 0.19$) \cite{EJ74} based
on the assumption of vanishing polarized strange sea.  Newer
experiments of the SMC \cite{AAA94} and SLAC \cite{AAB93}
collaborations give a somewhat larger value for $S_{EJ}^p$.
The polarized deep inelastic scattering experiments at CERN
have shown that, when supplemented with some information from
semileptonic decays, only a small fraction of the proton spin is
carried by valence quarks.

In view of the importance of the meson cloud for the Gottfried Sum
Rule violation, one might expect that it also plays an important role
for the nucleon spin. Indeed, simple estimates within the one pion
exchange model seem to indicate that the meson cloud effects could
contribute to the so-called spin crisis.  Moreover, due to the close
connection of the Ellis-Jaffe Sum Rule with semileptonic decays of
octet baryons (see section VI) we also have to evaluate the
consequences of the meson cloud for semileptonic decays of octet
baryons. Modifications of the axial-vector coupling constants due to
the meson cloud can be quite important \cite{JM911,JM912,SH93} since
the corresponding axial-currents are not protected against
renormalization due to the mesonic cloud.

The analysis of the EMC experiment under the assumption of $SU(3)$
symmetry (!) indicates that there may be a relation between the spin
and strangeness content of the nucleon. The value for the polarized
strange quark content of the nucleon obtained from such an analysis
is $\Delta s = -0.19$ (newer SMC\cite{AAA94} and SLAC\cite{AAB93}
experiments indicate a somewhat smaller amount), which suggests
a revision of the simple view of the nucleon. In meson cloud
models the strangeness content of the nucleon is generated by virtual
fluctuations into strange mesons ($K$ or $K^*$) and strange baryons
($\Lambda$, $\Sigma$ or $\Sigma^*$).
We stress the fact that the 'experimental' result obtained for the
polarized strange quark content has been obtained under the
assumption of $SU(3)$ symmetry, which is approximately realized in
semileptonic decays \cite{R90}.  On the other hand, $SU(3)$ symmetry
is violated as far as the masses of hadrons are concerned and it is
also violated in the sea quark distributions of the proton, which we
discuss in the present paper.  In models which include $SU(3)$
symmetry breaking effects the conclusion about the polarized
strangeness content can be modified. On the other hand, the
approximate validity of $SU(3)$ symmetry for the semileptonic decays
of baryons has to be an important check for any baryonic model.

The discussion above shows that in order to understand the structure
of the nucleon (the same is true for other baryons) one has to treat
consistently the flavour and spin content. Because of the surprising
success of the simple Cabibbo model, the semileptonic decays have to
be treated simultaneously.

The pion, being the lightest meson was the first included in the
deep-inelastic scattering \cite{S72,ABK81}. It was suggested in
Ref.\cite{HSB91} to include also other mesons which had proven to be
important in the low-energy nucleon-nucleon and nucleon-hyperon
scattering.  Their effects in deep inelastic scattering are discussed
in more detail in Ref.\cite{SS93}.

It has been suggested recently to study the virtual pion component
in the nucleon in deep inelastic scattering experiments
at HERA \cite{HLNSS94}. The semi-inclusive
reaction $ep \rightarrow e'nX$ is presently being studied by
the ZEUS collaboration, which has installed a test forward
neutron calorimeter to complement its leading proton
spectrometer \cite{ZEUS94}. It is also worth mentioning that a recent
lattice QCD calculations gave evidence for the importance of pion
loop effects for nucleon properties \cite{CL93}.
The renewed interest in the virtual pions (mesons) in the nucleon,
especially in connection to deep inelastic scattering,
requires a better understanding of the methods used up to now
in the literature on that subject.

In the meson cloud model the nucleon is viewed as a quark core, termed
a bare nucleon, surrounded by the mesonic cloud.  The convolution
model seems to be the best tool to understand the structure of such a
composed object.  The internal consistency of the convolution model
has been recently a subject of discussion \cite{MSM92,Z92}. A special
emphasis has been put on satisfying sum rules \cite{MSM92,Z92,SS93}.
When standard $t$-dependent form factors \cite{MSM92} are used, both
number and momentum sum rules cannot be satisfied automatically.
While number sum rules can be satisfied by independent adjustment of
cut-off parameters of the vertex form factors for the diagrams with
off-shell mesons and baryons, momentum conservation is violated to a
degree dependent on the functional form of the form factor (see also
\cite{DT92}). It has been pointed out recently that special care is
required, as far as form factors are concerned, in order to conserve
both sum rules in a natural way \cite{Z92}.

For the first time in the literature we present useful complete set of
formulas for the helicity dependent splitting functions of the nucleon
(octet baryon) into pseudoscalar/vector meson --- octet/decuplet
baryon states, i.e. we include all particles which turned out to be
crucial in modern meson exchange models for nucleon-nucleon and
nucleon-hyperon scattering at low energies \cite{MHE87,HHS89}.  In
addition to unpolarized deep-inelastic scattering we are also
interested in possible effects of the meson cloud for the polarized
deep-inelastic scattering.  The proton helicity flow to the pionic
cloud has been calculated recently in the relativistic light-cone
perturbation theory including $N\pi N$ and $N\pi\Delta$ vertices
\cite{Z93}.  The analysis of unpolarized deep-inelastic scattering
suggests that also vector mesons ($\rho$, $\omega$) may play an
important role \cite{SS93,MT93}, provided rather hard form factors are
used.  The importance of vector mesons ($S=1$) for deep inelastic
scattering of polarized particles can be even greater.  Therefore we
extend the formalism of Ref.\cite{Z93} to vector mesons.  We derive
useful formulas for both unpolarized and polarized deep-inelastic
scattering.

In comparison to our earlier analysis \cite{SS93}, where only
unpolarized deep-inelastic scattering has been considered, in the
present paper we analyze possible effects of the meson cloud also for
the polarized deep-inelastic scattering as well as for the
semileptonic decays of the octet baryons.  There are a few technical
details in which the present analysis differs from the previous one
\cite{SS93}. In Ref.\cite{SS93} we have assumed for simplicity a
dipole form of the vertex form factors with an universal cut-off
parameter for all mesons.  The value of the cut-off parameters of the
vertex form factors had been adjusted in Ref.\cite{SS93} to reproduce
the experimental data for $(\bar u(x) + \bar d(x))/2 -\bar
s(x)$. Since the universality of the dipole form factor is not obvious
and the fit to the $(\bar u(x) + \bar d(x))/2 -\bar s(x)$ data is not
very sensitive to the value of the cut-off parameter in the present
paper we take slightly different attitude.  Following \cite{AG81,Z92},
we estimate the free parameters of our model by analyzing high-energy
$pp\to nX$ and $pp\to\Delta^{++}X$ production processes in a
one-boson-exchange model.  In distinction to \cite{Z92} we include
also the effects of the vector meson Fock components. Then we first
discuss various effects of the meson--baryon Fock components on the
flavour structure of the nucleon and calculate the Gottfried Sum
Rule. Next we present the results of calculations for the axial-vector
coupling constants $g_A$ for all possible semileptonic decays of the
octet baryons. Finally, we calculate the Ellis-Jaffe Sum Rule for the
proton and neutron as well as the Bjorken Sum Rule.

\section{The Convolution Model}

We expand the nucleon wave function in terms of a few principle Fock
components. In this model the nucleon can be viewed as a bare nucleon
(core), surrounded by a mesonic cloud. The wave function of the
nucleon with helicity $+1/2$ can be schematically written as
\begin{equation}
|N{\mathord\uparrow}\rangle
=\sqrt Z\left[|N{\mathord\uparrow}\rangle_{bare}
+\sum_{BM}\sum_{\lambda\lambda'}
\int dy d^2 k_\perp
\phi_{BM}^{\lambda\lambda'}(y, k^2_\perp)
|B^\lambda(y,\vec k_\perp);M^{\lambda'}(1-y,-\vec k_\perp)\rangle
\right],
\label{Fock}
\end{equation}
where $\sqrt Z\phi^{\lambda\lambda'}_{BM}(y, k^2_\perp)$ is the
probability amplitude that a physical nucleon with helicity $+1/2$ is
in a state consisting of a virtual baryon $B$ with longitudinal
momentum fraction $y$, transverse momentum $\vec k_\perp$, and
helicity $\lambda$, and a virtual meson $M$ with momentum fraction
$1-y$, transverse momentum $-\vec k_\perp$, and helicity
$\lambda'$. Note, that the helicities $\lambda$ and $\lambda'$ need
not to add up to $+1/2$, since an additional relative angular momentum
between the particles is possible.  In Eq.(\ref{Fock}) $Z$ is the
standard wave function renormalization constant, which can be
interpreted as the probability of the bare nucleon \cite{DLY70}.

It can be expected, that the structure of the core is rather simple.
Presumably, it can be described as a three quark system in the static
limit. Of course, in the deep inelastic regime at higher $Q^2$
additional sea of perturbative nature is created unavoidably by the
standard QCD evolution.

The main idea of the convolution approach is that there are no
interactions among the particles in a multi-particle Fock state
during the interaction with the hard photon in deep inelastic
scattering (for a detailed discussion see \cite{MSM92}). This enables
one to relate the contribution of a Fock state $BM$ to the
nucleon structure function $F_2$, to the structure functions of
either the struck meson $M$ or the struck baryon $B$ (see Fig.~1a,b)
\begin{eqnarray}
\delta_{M} F_2^N(x)&=&\int_x^1 dy\> f_{MB/N}(y)
F_2^M\left({x \over y}\right),
\\
\delta_{B} F_2^N(x)&=&\int_x^1 dy\> f_{BM/N}(y)
F_2^B\left({x \over y}\right).
\end{eqnarray}
The convolution formulas can be written in an equivalent way
in terms of quark distributions:
\begin{equation}
q_N(x)=Z\left[q_{N,bare}(x)+
  \int_x^1 f_{MB/N}(y) q_M\left({x \over y}\right) {dy\over y}
 +\int_x^1 f_{BM/N}(y) q_B\left({x \over y}\right) {dy\over y} 
\right].
\label{vconv}
\end{equation}
The main ingredients in the formulas above are the {\em splitting
functions} $f_{MB/N}(y)$ and $f_{BM/N}(y)$, which are related to the
probability amplitudes $\phi_{BM}$
\begin{eqnarray}
f_{BM/N}(y)&=&
\int_0^\infty
dk^2_\perp\,\sum_{\lambda\lambda'}\,|\phi_{BM}^{\lambda\lambda'}
(y,k^2_\perp)|^2,
\label{fBM/N} \\
f_{MB/N}(y)&=&
\int_0^\infty
dk^2_\perp\,\sum_{\lambda\lambda'}\,|\phi_{BM}^{\lambda\lambda'}
(1-y,k^2_\perp)|^2.
\label{fMB/N}
\end{eqnarray}
Because the description of the nucleon as a sum of $MB$ Fock states
is independent of the reaction mechanism, the relation
\begin{equation}
f_{MB/N}(y)=f_{BM/N}(1-y)
\label{sym}
\end{equation}
must hold. It simply expresses the fact, that if a meson $M$ carries
a longitudinal momentum fraction $y$ of the nucleon momentum, the
remaining part of the nucleon is a baryon with the remaining
longitudinal momentum fraction $1-y$. Moreover this relation
automatically ensures global charge conservation
\begin{equation}
\langle f_{MB/N} \rangle=\langle f_{BM/N} \rangle
\end{equation}
and momentum conservation
\begin{equation}
\langle xf_{MB/N} \rangle+\langle xf_{BM/N} \rangle
=\langle f_{BM/N} \rangle,
\end{equation}
where $\langle f \rangle$ and $\langle xf \rangle$ are the first and
second moments of the splitting functions.

In this notation, the wave function renormalization constant is given
as
\begin{equation}
Z=\left[1+\sum_{MB}\langle f_{BM/N} \rangle\right]^{-1}.
\end{equation}

Both the Ellis-Jaffe Sum Rule and the axial-vector coupling constants
for semileptonic decays can be expressed with the help of
matrix elements of the axial-currents
$A^a_\mu=\bar q\gamma_\mu\gamma_5 (\lambda^a/2)q$ (see section VI),
where $\lambda^a$ are the
Gell-Mann matrices.
The expansion (\ref{Fock}) implies (see Fig.~2)
\begin{eqnarray}
\langle N|A^a_\mu|N\rangle_{dressed}
=Z\Big(\langle N|A^a_\mu|N\rangle_{bare}
&+&\sum_{B_1B_2M} \Delta f_{(B_1B_2)M/N}
  \langle B_1|A^a_\mu|B_2\rangle
\nonumber\\
&+&\sum_{M_1M_2B} \Delta f_{(M_1M_2)B/N}
  \langle M_1|A^a_\mu|M_2\rangle
\Big),
\label{axial}
\end{eqnarray}
where
\begin{eqnarray}
\Delta f_{(B_1B_2)M/N}&=&
\sum_{\lambda\lambda'} 2\lambda
   \int_0^1 dy \int_0^\infty d k^2_\perp\>
     \phi^{\lambda\lambda'}_{B_1M}(y, k^2_\perp)
     \phi^{*\lambda\lambda'}_{B_2M}(y, k^2_\perp) , \\
\Delta f_{(M_1M_2)B/N}&=&
\sum_{\lambda\lambda'} 2\lambda'
   \int_0^1 dy \int_0^\infty d k^2_\perp\>
      \phi^{\lambda\lambda'}_{BM_1}(1-y, k^2_\perp)
      \phi^{*\lambda\lambda'}_{BM_2}(1-y, k^2_\perp).
\end{eqnarray}

%%%%%
The essential ingredients in our model are the amplitudes
$\phi^{\lambda\lambda'}_{BM}(y,k^2_\perp)$.  To calculate these
quantities we employ the time-ordered perturbation theory (TOPT)
\cite{W66,S,DLY70} which has the advantage that the intermediate Fock
states can be written down explicitly:
\begin{equation}
\phi_{BM}^{\lambda\lambda'}
(\vec p,\vec k,\vec q=\vec p-\vec k)
= N_N{N_B \over (2\pi)^{3\over2}} {N_M \over (2\pi)^{3\over2}}
{V(\vec p,\uparrow;\vec k,\lambda;\vec q,\lambda') \over E_N-E_M-E_B}
\label{phi}
\end{equation}
(compare \cite{DLY70}).  This formula gives the amplitude of finding a
nucleon with momentum $\vec p$ and helicity $+1/2$ in a Fock state
where the baryon $B$ has the momentum $\vec k$, helicity $\lambda$ and
the meson $M$ the momentum $\vec q=\vec p-\vec k$ and helicity
$\lambda'$.  The factors $N_N$ ($N_B$) are the usual fermion wave
function normalization factors $N_N=\sqrt{m_N / E_N}$, $N_B=\sqrt{m_B
/ E_B}$; $N_M$ is a bosonic normalization factor $N_M=1/\sqrt{2 E_M}$.
The important feature of TOPT is, in contrast to the covariant
perturbation theory, that the intermediate particles are on their
mass-shell. Therefore, the vertex function $V$ in Eq.(\ref{phi}) can
be calculated by using on-mass shell spinors.  $V$ depends on a
particular model, i.e. on the form of the lagrangian used. In general
$V$ can be written as
\begin{equation}
V(\vec p,\vec k,\vec q)=\bar u_N(\vec p)_\alpha
v^{\alpha\beta\gamma} \chi_\beta(\vec q) \psi_\gamma(\vec k),
\end{equation}
where summing and averaging over all possible spin-states is
implicitly assumed. $\alpha$, $\beta$ and $\gamma$ are bi-spinor
and/or vector indices dependent on the representation used for
particles of a given type.  $\chi$ and $\psi$ are the wave functions
(field operators) of the intermediate meson and baryon,
respectively.

It has been shown \cite{DLY70} for the $\pi N$ case, that in the
infinite momentum frame (IMF) contributions of Fock states with
anti-particles vanish and only contributions with forward moving
particles survive.  This statement is also true for other Fock
states we are dealing with.  In the IMF-limit the momenta of the
particles involved can be parameterized in terms of $y$ and
${\vec k_\perp}$:
\begin{equation}
\vec k=y\vec p +{\vec k_\perp}, \quad
{\vec k_\perp}\cdot\vec p=0,\qquad \vec q=(1-y)\vec p-{\vec k_\perp}.
\label{kinIMF}
\end{equation}
In the limit $p=|\vec p|\to\infty$ only states with $y\in [0,1]$ do
not vanish \cite{DLY70}, and the amplitudes $\phi_{BM}$ can be
expressed as
\begin{equation}
\phi_{BM}(y,k^2_\perp)={1\over 2\pi \sqrt{y(1-y)}}
     {\sqrt{m_Nm_B}\>V_{IMF}(y,k^2_\perp)\over
           m_N^2-M^2_{BM}(y,k^2_\perp)}
\label{paIMF}
\end{equation}
with $M^2_{BM}$ being the invariant mass squared of the $BM$ Fock
state
\begin{equation}
M^2_{BM}(y,k^2_\perp)={m_B^2+k^2_\perp \over y}
                     +{m_M^2+k^2_\perp \over 1-y}.
\end{equation}
$V_{IMF}$ is the vertex-function in the IMF. In the formula above an
extra factor $(\pi p)^{-1/2}$ has been taken out.  It would cancel
when going to probability densities by an appropriate factor of the
jacobian of the transformation (\ref{kinIMF}).

%%%%%%%%%%%%%%%%%
If the vertex function used does not contain a derivative of the meson
field, we could have equally well used covariant perturbation theory
instead of TOPT. But if the vertex function used contains a derivative
of the meson field this is no longer true. To illustrate this point
let us consider the example of the pseudovector $N\pi N$ vertex, given
by
\begin{equation}
{\cal L}_{pv}=\bar u{\gamma_5}\gamma^\mu \partial_\mu\pi u,
\end{equation}
where for simplicity the coupling constant and isospin structure have
been suppressed. The standard covariant technique \cite{Z92,MSM92}
leads to the following splitting function of the meson
\begin{eqnarray}
f^{pv}_{MB/N}(y)&= & {1\over 16\pi^2} {1\over(1-y)^2 y}\int_0^\infty
dk^2_\perp |G(1-y,k^2_\perp)|^2 \nonumber \\ &&\cdot{(m_B+m_N)^2
[(m_N(1-y)-m_B)^2+k^2_\perp] \over [m_N^2-M_{BM}^2(1-y,k^2_\perp)]^2}.
\label{pvMB}
\end{eqnarray}
The result for the baryon is:
\begin{eqnarray}
f^{pv}_{BM/N}(y)&=& {1\over 16\pi^2} {1\over(1-y) y^2}\int_0^\infty
dk^2_\perp |G(y,k^2_\perp)|^2 \nonumber\\ &&
\Bigl(m_N^2m_B^2(1-y)^2+k^2_\perp(m_B+m_N)^2
     -2m_Nm_B(ym_M^2+k^2_\perp) \nonumber\\
&& \qquad +{1\over(1-y)^2}(ym_M+k^2_\perp)^2\Bigr)
{1\over[m_N^2-M^2_{BM}(y,k^2_\perp)]^2} .
\label{pvBM}
\end{eqnarray}
Here the two results are not related by Eq.(\ref{sym}), which leads
to violation of the charge-- and momentum conservation.  The reason
for this puzzle is, that by using a derivative coupling, an
additional off-shell dependence is introduced into the vertex
function, which cannot be suppressed in the IMF-limit. A way out is
to use TOPT.  Here, however, the problem arises how to choose the
meson energy in the vertex.  In principle, there are two possible
prescriptions:
\begin{itemize}
\item[A)] One uses in the vertex the meson four-momentum $q^\mu$:
$\bar u_N
{\gamma_5}\gamma_\mu (-i)q^\mu u_B$, i.e. the meson energy in the
vertex is $E_M$.  With this form of the vertex one
reproduces the baryon splitting function given by Eq.(\ref{pvBM}).
The meson splitting function is related to this result by
Eq.(\ref{sym}).

\item[B)] Instead of $q^\mu$ one uses the difference of the baryon
four-momenta $p^\mu-k^\mu$: $\bar u_N(p)
{\gamma_5}\gamma_\mu(-i)(p-k)^\mu u_B(k)$, i.e. the meson energy in
the vertex is $E_N-E_B$.  With this prescription one gets the
splitting function given by Eq.(\ref{pvMB}). Also here the resulting
splitting functions for the baryon and meson fulfill Eq.(\ref{sym}).
\end{itemize}
Thus TOPT, in contrast to a covariant calculation, is consistent
with the convolution approach. The remaining point to clarify is
which of the two prescription one should use. We will always use
prescription B, because in this prescription the splitting functions
for the pseudovector case are identical to those of the pseudoscalar
case, if the coupling constants are identified properly.  Moreover, in
this prescription the structure of the vertex is due to the baryonic
current only \cite{Tpriv}.

%%%%%%%%%%%%%%%%%%%%%%%%
\section{Form Factors}

The most general structure of the hadronic tensor must include the
dependence on off-shell effects \cite{MST94}. Because a realistic
calculation of such off-shell effects is not possible at present, we
shall neglect such effects in this paper.  Because of the extended
nature of the hadrons involved one has to introduce phenomenological
vertex form factors, which parameterize complicated unknown
microscopic effects.  The form factors are included in the calculation
by replacing the vertex function $V(y,k^2_\perp)$ by
$V'(y,k^2_\perp)=G(y,k^2_\perp)V(y,k^2_\perp)$. Now relation
(\ref{sym}), which can be shown to hold exactly in the case of
point-like particles (in either covariant and TOPT calculations),
imposes a severe restriction to these form factors:
\begin{equation}
G_{BM}(y,k^2_\perp)=G_{MB}(1-y,k^2_\perp).
\label{formsym}
\end{equation}
The form factors often used in meson exchange models and convolution
models are functions of $t$ only, the four-momentum squared of the
meson. It is to be expected that form factors which depend only on the
kinematical variables of only one of the two particles of the two-body
system, like the often used dipole form factor
\begin{equation}
G(t)=\left({\Lambda^2-m_M^2\over \Lambda^2-t}\right)^2,
\label{dipole}
\end{equation}
will not satisfy Eq.(\ref{sym}).
It can be shown \cite{MSM92} that these form factors do not conserve
basic quantities like charge and momentum simultaneously.  One
simple method to obtain form factors with the right symmetry is to
multiply a $t$-dependent form factor by a $u$-dependent one with the
same functional form with $m_M$ replaced by $m_B$
\begin{equation}
G_{sym}(t,u) = G(t,m_M)G(u,m_B).
\end{equation}
In terms of the IMF variables $y$ and $k^2_\perp$, $t$ and $u$, the
four momentum squared of the intermediate baryon, is given by
\begin{eqnarray}
&&t= -{k^2_\perp\over y}-(1-y)({m_B^2\over y}-m_N^2),
\label{deft} \\
&&u= -{k^2_\perp\over 1-y}-y({m_M^2\over 1-y}-m_N^2).
\label{defu}
\end{eqnarray}

The importance of using such symmetric form factors was noticed only
recently \cite{Z92}. Another possible approach, to fix the
cut-off parameters to assure number sum rules (global charge
conservation) (see \cite{MSM92,DT92}) is in our opinion somewhat
arbitrary, and does not guarantee momentum conservation.

In numerical calculations we use vertex form factors in the
exponential form
\begin{equation}
G_{BM}(y, k^2_\perp)
=exp\Big[{ m_N^2-M^2_{BM}(y, k^2_\perp)\over 2\Lambda^2}\Big],
\label{form}
\end{equation}
where $M^2_{BM}(y, k^2_\perp)$ is the invariant mass squared of the
baryon-meson $BM$ Fock state
\begin{equation}
M^2_{BM}(y, k^2_\perp)
={ m_B^2+ k^2_\perp\over y}+{ m_M^2+ k^2_\perp\over 1-y}.
\end{equation}
A form factor of this type, first introduced in Ref.\cite{Z92},
fulfills the necessary symmetry conditions Eq.~(\ref{sym}).
Parameterization of the vertex form factors in terms of the invariant
mass of the two-body system is very natural in the light-cone
approach.  Furthermore, the light-cone approach is the best suited for
applications to deep inelastic scattering.

In order to fix the cut-off parameters $\Lambda$ of the form factors
we use high-energy baryon production data. The neutron production
data in the $pp\to nX$ scattering seems to be best tailored for
extracting the cut-off parameter for the $N\pi$ and $N\rho$ Fock
states. Because we shall limit ourselves to data for relatively low
exchanged four-momenta, it is reasonable to assume that the
neutron is produced by a simple one-boson-exchange mechanism (OBE)
(see Fig.3). In this model the differential cross section can be
described as a product of the probability for a proton being in a
$n\pi^+$ or $n\rho^+$ Fock state and the total cross section of
$\pi^+(\rho^+)p$ scattering. In general, the invariant cross section
for $pp\to BX$ production in the OBE model is
\begin{equation}
E{d^3\sigma(pp\to BX)\over d^3p}
={y\over\pi}{d^2\sigma\over dyd k^2_\perp}
={y\over\pi}\sum_{\lambda\lambda'}
  |\phi_{BM}^{\lambda\lambda'}(y, k^2_\perp)|^2
  \cdot \sigma_{tot}^{Mp}(s(1-y)).
\label{cross}
\end{equation}
Here $y$ is the longitudinal momentum fraction of the baryon with
respect to the momentum of the incoming proton; $\vec k_\perp$ the
corresponding transverse momentum.  Specialized to neutron production
with $\pi^+$-exchange this formula reads:
\begin{eqnarray}
E{d^3\sigma(pp\to nX)\over d^3p}
&=&{g^2_{pn\pi^+}\over 16\pi^3}{(1-y)^2 m_N^2+ k^2_\perp\over
[ m_N^2-M^2_{N\pi}(y, k^2_\perp)]^2}
{|G_{NN\pi}(y, k^2_\perp)|^2\over y(1-y)} \\
&&\cdot\sigma^{\pi p}_{tot}(s(1-y)). \nonumber
\end{eqnarray}
For $\rho^+$-exchange the formula is much longer and will be not
presented here. The necessary ingredients can be found in Appendix B.

These formulas are strictly valid only in the IMF-limit, but for a
sufficiently high energy they are a good approximation. In the
following we neglect the slow energy dependence of the total $\pi N$
cross section and use a value of $\sigma^{\pi^+ p}_{tot}=23.8\pm
0.1 mb$ \cite{P92}. In addition, due to the lack of experimental
data we also assume $\sigma^{\rho^+p}_{tot}=\sigma_{tot}^{\pi^+p}$.

In Fig.4 we show a fit by the OBE model to the experimental data
(the $ k^2_\perp=0$ data are taken from Ref. \cite{FM76} with
$\sqrt{s}=53\>GeV$; the other data are taken from Ref. \cite{B178}
with $p_{LAB}=24\ GeV/c$).  In this calculation we take
$g^2_{pp\pi^0}/4\pi=13.6\pm0.1$ \cite{TRS91} and
$g^2_{pp\rho^0}/4\pi=0.84$, $f_{pp\rho^0}/g_{pp\rho^0}=6.1$
\cite{HHS89}. As a criterion for the fit we have assumed that the
calculated result must not exceed the experimental data. For low $
k^2_\perp$ $\pi$-exchange is the dominant contribution, the
$\rho$-exchange plays a rather marginal role. For higher $
k^2_\perp$ $\rho$-exchange becomes the dominant mechanism.  Thus the
cut-off parameter for the $NN\pi$ and $NN\rho$ vertices can be fixed
almost unambiguously. The fit to the experimental data yields
$\Lambda_{NN\pi}=\Lambda_{NN\rho}=1.10\pm0.05\>GeV$.

We obtain a good description of the experimental data at
intermediate $y=0.4$--$0.9$. Outside this region the fit fails. This
is to be expected, because for low $y$ one could expect additional
contributions from e.g. multi-meson exchange processes.  For our
purpose (later we shall concentrate on moments of $x$-distributions
rather than the distributions themselves) the assumption of a OBE
mechanism is sufficient, as the maximum cross section lies at
intermediate $y$, where the mesons can be treated as not reggeized.
The correct $ k^2_\perp$-dependence obtained from the fit is a
further argument in favor of the simple OBE model. Despite the
success of the fit one should bear in mind that we obtain only an
upper bound of the $\rho$-exchange contribution.

In order to extract information about the $NK\Lambda$ vertex, one can
examine high-energy $\Lambda$ production in the $pp\to\Lambda X$
reaction in the same way as neutron production.  Also here we
assume that the dominant mechanism is a OBE mechanism, $K$ or
$K^*$-exchange. The differential cross section for
$K$-exchange is given by
\begin{eqnarray}
E{d^3\sigma(pp\to\Lambda X)\over d^3 p}&=&
g^2_{p\Lambda K^+}\over 16\pi^3}{(y m_N-m_\Lambda)^2+ k^2_\perp\over
[ m_N^2-M^2_{\Lambda K}(y, k^2_\perp)]^2}
{|G_{N\Lambda K}(y, k^2_\perp)|^2\over y(1-y)\\
&&\cdot\sigma^{Kp}_{tot}(s(1-y)). \nonumber
\end{eqnarray}
For $K^*$-exchange we refer to Eq. (\ref{cross}) and Appendix B.

We have taken $g^2_{p\Lambda K^+}/4\pi=14.7$, which is related by
$SU(6)$ symmetry to the $g^2_{pp\pi^0}$ given before. The coupling
constant obtained from experiment is $15.4\pm1.5$ \cite{TRS92} which
supports the assumption of $SU(6)$ symmetry. Also for the coupling
constants involving vector mesons we assume $SU(6)$ symmetry (for
details see \cite{HHS89}). The total cross section
$\sigma_{tot}^{K^+p}$ is $19.9\pm0.1mb$ \cite{P92}. For
$\sigma_{tot}^{K^{*+}p}$ we assume the same value.  The fit of the
cut-off parameters $\Lambda_{N\Lambda K}$ and $\Lambda_{N\Lambda
K^*}$ is shown in Fig.4.  The general features of the fit are similar
to those found for neutron production. Also, the cut-off parameters
$\Lambda_{N\Lambda K}=\Lambda_{N\Lambda K^*}=1.05\pm0.05\>GeV$
obtained from the fit are quite similar to those found for $\pi$- and
$\rho$-exchange. Thus, it is possible to assume a kind of
universality and use a cut-off parameter $\Lambda=1.08\>GeV$ for
all vertices involving octet baryons and pseudoscalar or vector
mesons. This is rather attractive, that the same cut-off parameters
can be used for pseudoscalar and vector mesons.

In the literature, instead of the symmetric form factor given by
Eq.(\ref{form}), a dipole form factor Eq.~(\ref{dipole}) is often
used.  As discussed above, the use of such form factors leads to a
violation of basic symmetries. In Fig.5 we show, for example, the
differential cross section for neutron production calculated using the
dipole form factor ($\Lambda=1.2\>GeV$ \cite{SS93}). As seen from the
figure, one gets a satisfactory description for small $
k_\perp^2$. One cannot, however, describe the data at large $
k^2_\perp$. Comparison of Fig.4 and Fig.5 clearly indicates a
preference for the symmetric form factor (Eq. \ref{form}).

The violation of the Gottfried Sum Rule is an interplay between the
$\pi N$ and $\pi \Delta$ components and crucially depends on vertex
form factors, i.e. its functional form and cut-off parameter.  For
instance, taking the dipole form and assuming the same cut-off for
both components leads to strong cancellations. As a consequence only
about half of the experimentally observed violation can be explained
\cite{SS93,K191,K291}.  In order to shed light on the $N\pi\Delta$
vertex we follow \cite{AG81} and examine $\Delta^{++}$ production in
$pp$ and $p\bar p$ collisions. The differential cross section for
one-pion-exchange (Fig.6) is given by:
\begin{eqnarray}
E{d^3\sigma(hp\to \Delta^{++}X)\over d^3p}&=&
{g^2_{p\Delta^{++}\pi^-}\over16\pi^3}
{|G_{N\Delta\pi}(y,k^2_\perp)|^2\over y(1-y)}  \nonumber\\
&&\cdot{[(y m_N+m_\Delta)^2+k^2_\perp]^2
        [(y m_N-m_\Delta)^2+k^2_\perp]\over
   6y^2 m_\Delta^2[ m_N^2-M^2_{\Delta\pi}(y,k^2_\perp)]^2}
\label{delta}  \\
&&\cdot\sigma^{\pi h}_{tot}(s(1-y)). \nonumber
\end{eqnarray}
The differential cross section for $\rho$-exchange can be calculated
analogously (see Eq. \ref{cross} and Appendix B).

As coupling constants we take
$f^2_{p\Delta^{++}\pi^-}/4\pi=12.3\>GeV^{-2}$ and
$f^2_{p\Delta^{++}\rho^-}/4\pi=34.7\>GeV^{-2}$ \cite{HHS89}.  The
available $\Delta^{++}$-production spectra \cite{B179} are biased by
an ambiguity in extraction of background contributions.  Thus, in
Fig.7 two scales are shown (different background assumptions). The
experimental data shown in Fig.7 have been taken at three different
energies $\sqrt{s}=100$, $200$ and $360\>GeV$. We do not distinguish
the data taken at different energies, as no energy dependence is
visible. This fact is in agreement with the simple assumption of an
OBE mechanism and supports the IMF-approach. As far as the absolute
normalization is concerned some extra experimental information is
needed. We use more precise $y$-integrated-spectra dependent only on
$k^2_\perp$ \cite{B179,B180} for this purpose.  The result of the fit
shown in Fig.7 and Fig.8 yields
$\Lambda_{N\Delta\pi}=\Lambda_{N\Delta\rho}=0.98\pm0.05\>GeV$. In
contrast to the $n$ and $\Lambda$ production an unambiguous
distinction between the $\pi$ and $\rho$ exchange contributions is
not possible. For simplicity, we have assumed the same cut-off
parameters for both, in analogy to the octet baryon
production.

We assume that the cut-off parameters for the whole decuplet to be
the same as those obtained from the $pp\to\Delta^{++}X$ production
process.  This universality within an $SU(3)$ multiplet greatly
reduces the number of, in principle unknown, parameters.

To reduce the number of parameters further, we employ $SU(6)$
symmetry as mentioned above to relate coupling constants. Thus only
the coupling of the $\omega$ to the nucleon remains to be fixed. It
has to be treated differently, because the $\omega$ is an almost
ideal mixture of an octet and a singlet \cite{HHS89}
\begin{equation}
\omega={1\over\sqrt 3}\omega_8+\sqrt{2\over 3}\omega_1.
\end{equation}
The $\omega$-couplings were found to be: $g^2_{pp\omega}/4\pi=8.1$
and $f_{pp\omega}/g_{pp\omega}=0$ \cite{HHS89}.

It is interesting to ask the question of how sensitive is our model
to the functional form of the form factor. In the following we
restrict ourselves to form factors which have the correct symmetry.
The simplest are the symmetric monopole ($n=1$) or
dipole ($n=2$) form factors:
\begin{equation}
G_n(t,u)=\left(\Lambda^2- m_M^2\over\Lambda^2-t\right)^n
         \left(\Lambda^2- m_B^2\over\Lambda^2-u\right)^n .
\label{symform}
\end{equation}
In principle, one might expect that they produce results different
from the exponential form (Eq. \ref{form}) for the first moments of
the splitting functions, because they cut off the integrands much
more softly.  Surprisingly, this is not the case. In Table 1 we show
values of cut-off parameters of the exponential, monopole and dipole
form factors fitted to the neutron production data and the
corresponding moments of the splitting functions. Almost the same
first moments are obtained with the different functional forms. Our
results are almost insensitive to the form of the form factors used;
it is rather the symmetry (Eq. \ref{sym}) which allows one to
describe the data in a broad range of $y$ and $ k^2_\perp$.

\section{Meson Cloud Effects on the Flavour Structure of the Nucleon}

The renewed interest \cite{HM90,K191,K291,SST91,HSB91,SS93} in the
meson cloud of the nucleon comes from the fact that it provides a
natural
explanation of the $\overline{d}$-$\overline{u}$ asymmetry and, as a
consequence, the violation of the Gottfried Sum Rule. The Gottfried
Sum Rule can be expressed as
\begin{equation}
S_G=\int_0^1 [F_2^p(x)-F_2^n(x)]{dx\over x}
   ={1\over3}\int [u(x)+\bar u(x)-d(x)-\bar d(x) ]\; dx
   ={1\over 3}+{2\over 3}A_{\bar u-\bar d},
\label{gottfried}
\end{equation}
where $q$ ($\bar q$) are the light quark (antiquark) distributions
in the proton and $A_{\bar u-\bar d}=\smallint_0^1 [\bar u(x)-\bar
d(x)]\,dx$.
In obtaining Eq.(\ref{gottfried}) isospin symmetry within the nucleon
doublet has been assumed\footnote{The violation of the isospin
symmetry of proton and neutron could in principle be another
important effect \cite{MSG93}. Up to now no quantitative estimate of
this effect exists.}.
In accordance with the simple structure of the bare nucleon,
it is natural to expect that the only sea in the bare nucleon is of
perturbative origin. Such a sea, being $SU(2)$ symmetric, does not
contribute to $A_{\bar u-\bar d}$.
In our model the asymmetry is
caused by the valence quarks in the virtual mesons. Only $\pi$ and
$\rho$ mesons contribute to the asymmetry \cite{SS93}, which can be
expressed in terms of the first moments of the corresponding
splitting functions
\begin{equation}
A_{\bar u-\bar d}=  Z\Big[
  -\langle f_{\pi^+n/p} \rangle +\langle f_{\pi^-\Delta^{++}/p}\rangle
  -\langle f_{\pi^+\Delta^0/p}\rangle
  -\langle f_{\rho^+n/p} \rangle +\langle f_{\rho^-\Delta^{++}/p}
\rangle
  -\langle f_{\rho^+\Delta^0/p} \rangle
   \Big].
\end{equation}
Contributions of a perturbative nature as well as any symmetric sea
in mesons and/or baryons would not change this result, because $S_G$
is sensitive only to the difference $\bar u-\bar d$
(Eq. \ref{gottfried}). In Table 2 we show various contributions to
the $\bar u -\bar d$ asymmetry and to the Gottfried Sum Rule. The
inclusion of the $\pi N$ Fock component leads to a reduction of $S_G$
from $1/3$ to $0.230$.
In comparison to earlier works where an universal cut-off
parameter of the dipole form factor has been assumed for simplicity
\cite{K291,SS93}, the procedure proposed in section III leads to
important
modifications. We find a much smaller probability of the $\pi\Delta$
(6.2 \% in comparison to 16.1 \% in Ref.\cite{SS93})
Fock component and a substantially larger probability of the $\rho N$
component (10.9 \% in
comparison to only 1.6 \% in Ref.\cite{SS93}).
Summarizing, including all components in the
nucleon wave function expansion (\ref{Fock}) leads to a good agreement
with the NMC result \cite{A91} of the Gottfried Sum Rule $S_G=0.24\pm
0.016$ (see Table 2).
It should be noted here, that the 'true' experimental value
of $S_G$ might be even somewhat smaller than the NMC result due to
shadowing effects in the deuteron \cite{Z92,MT93S}.

In calculating the $\bar u(x) - \bar d(x)$ asymmetry in the present
paper we have neglected the interference diagrams like $\pi^{0} -
\eta$ or $\rho - \omega$.  While the contributions of such diagrams to
the number and momentum sum rule cancel, their contribution to $\bar
u(x)$, $\bar d(x)$ as well as to the $\bar u(x) - \bar d(x)$ asymmetry
is not protected by any conserved current and is in principle
possible.  A reliable calculation of the $x$-dependence of the $\bar u
- \bar d$ difference is rather difficult and requires microscopic
models for the mesons involved. In contrast to the $x$-dependence the
bulk effect of $\bar u - \bar d$ can be easily estimated assuming that
the discussed mesons are members of the $SU(3)$ multiplets.  The
contribution of the interference diagrams of two mesons to the $\bar u
- \bar d$ difference can be calculated in a full analogy to the
diagonal terms:
\begin{equation}
\langle\bar u - \bar d\rangle = 
  \int_0^1 dy \int d k^2_{\perp} \sum_{\lambda\lambda'}
\phi_{B M_1}^{\lambda\lambda'}(y,k^2_\perp) 
\phi_{B M_2}^{\lambda\lambda',*}(y,k^2_\perp)
\langle M_2| \bar u-\bar d|M_1\rangle,
\label{interfer}
\end{equation}
where the matrix element $\langle M_2| \bar u-\bar d|M_1\rangle$ can
be calculated using the $SU(6)$ wave functions of the mesons involved.
We find $\langle\bar u - \bar d\rangle = 0.0050$ and $-0.0011$ for
$\pi^{0} - \eta$ and $\rho - \omega$, respectively. These
contributions have to be compared to $-0.12$ ($\pi$) and $-0.073$
($\rho$) obtained from the diagonal mesonic terms. There are different
reasons for the smallness of these contributions. While the $\pi^{0} -
\eta$ is small due to a small coupling constant $g_{NN\eta}$, the
$\rho - \omega$ contribution is small due to the vanishing tensor
coupling constant $f_{NN\omega}$. Thus the peculiarities of the
nucleon-meson-nucleon couplings allows one to safely neglect the
interference terms.

As has been emphasized in Ref.\cite{HSB91}, an attractive feature of
the meson cloud is that a large fraction of the nucleon sea can be
attributed to the virtual mesons (see also Ref.\cite{ABK81}). Let us
consider this issue in more detail in the case of the present
model. For this purpose the global CCFR experimental data \cite{F90}

\begin{equation}
\begin{array}{rll}
\kappa &\displaystyle{={2\langle x s\rangle
           \over \langle x \bar u\rangle+\langle x \bar d \rangle}}
       &=0.44^{+0.09+0.07}_{-0.07-0.02}, \\
\noalign{\vskip2mm}
\eta_s &\displaystyle{={2\langle x s\rangle
           \over \langle x u\rangle+\langle x d\rangle}  }
       &=0.057^{+0.010+0.007}_{-0.008-0.002}, \\
\noalign{\vskip2mm}
R_Q    &\displaystyle{
 ={\langle x \bar u\rangle+\langle x \bar d\rangle
    +\langle x \bar s\rangle
   \over\langle x u\rangle+\langle x d\rangle+\langle x s\rangle}}
     &=0.153\pm0.034 \\
\end{array}
\label{ratio}
\end{equation}
obtained at an average value of $Q^2=16.85\,(GeV/c)^2$ are useful.
In the formulas above $\langle x q\rangle$ denote second moments of
the quark distributions $q(x)$. Obtaining the quantities given by
Eq.(\ref{ratio}) in our model is not straightforward. First of all,
the experimental data has been measured at relatively large
$Q^2$. Therefore, the model has to be extended by the inclusion of
extra sea of a perturbative nature. Secondly, the quantities in
Eq.(\ref{ratio}) are sensitive to the valence quark distributions
which cannot be calculated fully consistently
\cite{SS93}. In spite of these difficulties it seems instructive to
make an estimate of those quantities. It is consistent
within our approximation scheme to assume a simple structure of bare
baryons and take into account fully dressed mesons.

Calculation of the CCFR quantities implies knowledge of the
$x$-dependences of the quark distributions. For the mesons we take
quark distribution functions known from the analysis of the Drell-Yan
processes in the pion-nucleus collisions \cite{B83,SMRS92}.  For the
bare baryons the situation is somewhat more complicated, and some
approximations are unavoidable. For the purpose of comparison with
the global quantities like those given by Eq.(\ref{ratio}) it is
sufficient to approximate valence quark distributions by those of the
physical baryons. A similar approximation for the sea quarks is,
however, not realistic as within our model a large fraction of the
nucleon's sea is included explicitly via the pionic (mesonic)
cloud. As discussed in Ref.\cite{SSG94}, this effect can be
approximately included by reducing the total nucleon's sea, known
from DIS, by a factor $R_{sea}$.

Our model, as discussed later, predicts both asymmetry between $\bar
u$ and $\bar d$ quark distributions and a suppression of the strange
sea in comparison to the non-strange sea.  On the other hand,
perturbative QCD predicts $\bar u$ - $\bar d$ symmetry and allows for
relative suppression of the strange component.  It is an empirical
fact that at $Q^2$ = 4 GeV$^2$ the strange sea is suppressed by a
factor 2.  It is interesting how much of this effect is of
perturbative/nonperturbative nature.  One could use the CCFR parameter
$\kappa$ (see Eq.(\ref{ratio})) to fix the strangeness suppression
factor $R_{str}$ for the sea quark distributions in mesons, baryons
and the bare nucleon (not dressed with mesons),
\begin{equation}
s_{sea}(x,Q^2) = \bar s_{sea}(x,Q^2) =
R_{str} \, \bar u_{sea}(x,Q^2) =
R_{str} \, \bar d_{sea}(x,Q^2) \; .
\end{equation}

In the present paper we adjust the two unknown parameters $R_{sea}$
and $R_{str}$ to the CCFR quark distributions $\bar q(x) = \bar u(x) +
\bar d(x) + \bar s(x)$ \cite{M92} and to the CCFR parameter $\kappa$
\cite{R93}.  In principle both $\bar q(x)$ and $\kappa$ depend
explicitly on $R_{sea}$, $R_{str}$. In practice the situation is
simpler because $\bar q(x)$ depends mainly on $R_{sea}$ and is almost
independent of $R_{str}$ while $\kappa$ depends mainly on $R_{str}$
and is almost independent of $R_{sea}$. This allows us to extract the
two parameters. If we take the $S_{0}'$ MSR \cite{MSR93} sea
parameterization for the bare nucleon (we assume that the valence
quark distributions in baryons are related via $SU(3)$ symmetry), we
find $R_{sea} = 0.4$ and $R_{str} = 0.75$.  The corresponding $\bar
q(x)$ is compared with the "experimental" sea quark distribution of
the CCFR collaboration \cite{M92} in Fig.9 and the resulting parameter
$\kappa$ = 0.40. The other CCFR parameters are then: $\eta_{s} =
0.069$, $R_{Q} = 0.201$, slightly above their experimental
counterparts. The small overestimation can be easily understood
because these two quantities are sensitive to the valence quark
distributions which cannot be calculated fully consistently
\cite{SS93}. We expect the values of $R_{Q}$ and $\eta_s$ to be biased
by the approximations made up to 15\%.

While $R_{sea}$ depends to some extend on the parameterization of the
sea quark distributions used, $R_{str}$ is fairly stable.  Therefore
the result of the fit suggests that the empirical suppression of the
perturbative strange sea component at $Q^2$ values of a few GeV$^2$
is about 0.75 in comparison to the total strange sea suppression of
0.5.  The difference is due to nonperturbative effects of the meson
cloud.

Not only global quantities like the Gottfried Sum Rule or the CCFR
parameters are of interest. Also the $x$-dependence of various
quantities is helpful in testing models.
It is known that sufficiently far above the strangeness production
threshold the QCD evolution equations \cite{AP77} lead to the
production of a perturbative $SU(3)$ symmetric sea.  In this context
it is useful to study differences of quark distributions instead of
the distributions themselves, especially those describing the
$x$-dependence of the $SU(2)$ and $SU(3)$ symmetry violation of the
sea quarks \cite{SS93,SSG94}.

The $x(\overline{d}(x) - \overline{u}(x))$ is the quantity which
describes the $x$-dependence of the $SU(2)$ symmetry breaking of the
nucleon sea. This quantity cannot be easily obtained from the
integrand of $S_G$ which, in addition to $\overline{d}-\overline{u}$,
involves the $x$-dependence of valence quark distributions (see
Eq.\ref{gottfried}), making the extraction of the interesting
quantity rather difficult.  The analysis of dilepton production in
$pp$ and $pd$ collisions seems to be much better in this respect
\cite{G92,SSG94}.

$x(\bar d - \bar u)$ is of special importance since it can be used to
verify our model as it does not require further parameters.  Having
fixed the cut-off parameters of the vertex form factors from hadronic
reactions it can be obtained parameter free. If one assumes that the
sea of the bare nucleon is $SU(2)$ symmetric, then only pions (mesons)
contribute to $x(\bar d - \bar u)$.  If, in addition, one assumes that
the pion sea is also $SU(2)$ symmetric (there are no reasons for an
asymmetry), then the whole effect comes from the valence quarks in the
pion which are relatively well known from the Drell-Yan processes.  In
Fig.~10a we show the predictions of our model for three different pion
structure functions: the leading order (LO) parameterization of the
NA3 collaboration \cite{B83} (dotted line), the next-to-leading order
(NLO) parameterization of Ref.\cite{SMRS92} (solid line) and structure
function of the pion calculated in a model of the radiatively
generated sea \cite{GRV92_pi} (dashed line).  As seen from the figure
the results for different parameterizations are rather similar.  In
Fig.~10b we compare our result (solid line) with parameterizations of
the quark distributions which have been fitted to the world data on
DIS and Drell-Yan processes. It is worth noting that our result is
almost identical for $x < 0.2$ with a recent NLO parameterization
MSR(A) of the MSR group \cite{MSR94} which includes both the new HERA
data \cite{H1,ZEUS} and the experimental result of the Drell-Yan NA51
experiment \cite{B94}. The Drell-Yan experiment planned at Fermilab
\cite{G92} will be a severe test of our model and should shed new
light on the origin of the Gottfried Sum Rule violation.

Let us consider now the $x$-dependence of $x({1 \over 2} (\bar u +
\bar d)-\bar s)$ which has been extracted by Kumano \cite{K191}
from the experimental data of the E615 Collaboration \cite{H89}. In
Fig.~11 we show the prediction of our model. In analogy to the $x(\bar
d - \bar u)$ difference we show separately the effect of the valence
quarks in mesons (dashed line). It explains already a significant
fraction of the total strange quark suppression.  The dotted line
shows in addition the effect when the extra perturbative
contributions, as described in this section, are included.  The
slight overestimation of this experimental data may suggest even
smaller suppression of the perturbative component with respect to
that obtained based on the CCFR data.

The net strangeness of the nucleon is zero, which requires: $\int s(x)
dx = \int \bar s(x)dx$. It has been customarily assumed that $s(x) =
\bar s(x)$. There is no basic principle that forces one to this
assumption other than the fact that it appears as a consequence of a
perturbative approach. In this context nonperturbative effects can be
of crucial importance. Recently \cite{BW92}, a possible charge
$x$-asymmetry of the strange sea has been studied in the chiral
Gross-Neveu model.  Although a large asymmetry has been found, no
absolute normalization of the effect was possible.  A similar analysis
has been performed somewhat earlier by Signal and Thomas
\cite{ST87}. In their approach the result strongly depends on the bag
radius.  At present, experimental proposals to investigate this
effect are under discussion \cite{vH}.  The presence of (strange
meson)-(strange baryon) Fock components in the nucleon wave function
suggests the asymmetry in a natural way, as $\overline{s}$ quarks are
constituents of 'light' mesons ($K, K^{*}$) and $s$ quarks are
constituents of 'heavy' baryons ($\Lambda, \Sigma, \Sigma^{*}$). Let
us estimate the effect in our model.  In distinction to
Ref.\cite{BW92}, we can estimate the absolute magnitude of the effect
as our free parameters have been determined (see the preceding
section) by fitting to the experimental data on $pp \to \Lambda X$
production at high energy.  The contributions from the valence strange
quarks in baryons and valence anti-strange quarks in mesons are shown
in Fig.~12a for two different sets of quark distributions: (a) LO
quark distributions in the pion \cite{B83} and in the nucleon
\cite{O91} (dashed line) and (b) NLO quark distributions in the pion
\cite{SMRS92} and in the nucleon \cite{MSR93} (solid line).

Making the plausible assumption that the effects discussed in the
present paper are the only source of the $s-\bar s$ asymmetry, the
$s-\bar s$ difference can be obtained without additional free
parameters.  Naively one could expect $s > \bar s$ in the large-$x$
region and $s < \bar s$ in the small $x$ region (see also
\cite{BW92}). We get a rather opposite effect to the naive
expectation as seen in Fig.~12b, where we show predictions of our model
using the same structure functions as in Fig.~12a. These
contributions are rather small in comparison to similar contributions
to $\bar u$ and $\bar d$ quark distributions (see for instance
Ref.\cite{SSG94}).  As seen from the figure, in comparison to the
$\bar d - \bar u$ difference, $s - \bar s$ depends not only on the
parton distributions in mesons but also on the parton distributions in
baryons. This difference can be expressed in terms of the
convolution integrals
\begin{eqnarray}
s(x) - \bar s(x)
= Z \Bigg[&&
\sum_{B} \int_{x}^{1} f_{BM/N}(y) \,
s_{val}^{B}({x\over y}) \, \frac{dy}{y}
\nonumber \\
-&&
\sum_{M} \int_{x}^{1} f_{MB/N}(y) \,
\bar s_{val}^{M}({x\over y}) \, \frac{dy}{y} \Bigg] \; .
\label{sas}
\end{eqnarray}
Eq. (\ref{sas}) shows that the shape of the $s - \bar s$
difference depends both on the splitting functions $f$ and valence
quark distributions both in $B$ ($s_{val}^B$) and $M$ ($\bar
s_{val}^M$). Therefore, the
final result depends on two competing effects: (a) $f_{BM/N}(y)$ peaks
at $y>0.5$ (for the $\Lambda K$ case, for instance, it is, however,
close to $y=1/2$); (b) the valence quarks in mesons are concentrated
at larger $x$ in comparison to the valence quarks in baryons (2 vs. 3
valence quarks, respectively).  It is the second effect which wins in
our case. Because the $s - \bar s$ difference depends on these two
competing effects it is somewhat less reliably determined in
comparison to the $\bar d - \bar u$ difference.

The total (anti)strange quark distributions obtained by applying
the procedure described above are shown in Fig.12c. In
addition we present a decomposition into contributions from
the meson's sea and that from the bare nucleon and bare baryons.

Can the strange quark density be obtained in a reliable way more
directly from experimental data?  It was proposed that the opposite
sign dimuon production in the charge current (anti)neutrino DIS can
be used as an unique probe of charm particle production and the
strange sea content of the nucleon \cite{R93}. The mass of the charm
quark introduces a threshold suppression into the dimuon production
rate which has been overcome in Ref.\cite{R93} by applying a simple
slow-rescaling model \cite{GP76}.  For comparison in Fig.~12c we show
also the result obtained by the CCFR collaboration in the E744 and
E770 experiments at Fermilab \cite{R93} (dashed line).

By parameterizing the sea quark distributions by the simple
$xq_{sea}(x) = A(1-x)^{\alpha}$ form, the CCFR collaboration has
found a quantitative indication that the strange sea in the nucleon
is softer than the non-strange sea with $\alpha$ = 9.45 vs. 6.95
\cite{R93} for the non-strange distributions.  This observation is
consistent with the prediction of our model.  In the large-$x$
($>$0.25) region the non-strange sea is completely dominated by the
mesonic (mainly pionic) effects which exhaust practically the total
strength obtained from the analysis of DIS data.  In contrast in the
strange sector the mesonic effects are strongly suppressed by the
meson/baryon mass effects.  Thus the model predicts less strength in
the large-$x$ region in comparison to the non-strange sea quark
distributions.

Although we get a rather reasonable agreement with the CCFR dimuon
data we do not consider it as an ultimate test of our model. First of
all, as already discussed, the mesonic effects are not the only
source of the strange quarks at $Q^2$ of a few GeV$^2$. Secondly, we
want to stress that the interpretation of the CCFR charm production
data should be taken with some grain of salt. It has been
demonstrated recently \cite{BGNPZ94} that a clear interpretation of
the CCFR result in terms of the strange quark distributions is
questionable at least in the case of the ($W^{\pm}$ -- gluon) fusion
component of the strange sea and the experimentally measured cross
sections may include also contributions of different origin.
Furthermore, it is not fully clear whether the slow-rescaling
procedure includes the mass effects in a correct way.  Our effect of
the $s - \bar s$ asymmetry is very small. Recently, this asymmetry
has been studied by the CCFR collaboration.  They have found a very
little
effect \cite{Baz94}, almost consistent with zero, in agreement with
our predictions.

\section{The spin of the proton}

In terms of quark distributions the Ellis-Jaffe Sum Rule can be
written as
\begin{equation}
S_{EJ}^p=\int_0^1 g^p_1(x)\>dx={1\over 2}\big({4\over9}\Delta u
+{1\over9}\Delta d+{1\over 9}\Delta s\big),
\end{equation}
with $\Delta q(x)$ being the polarized quark distributions
\begin{equation}
\Delta q = \int_{0}^{1}
[q^{\mathord\uparrow}(x)
 - q^{{\mathord\downarrow}}(x)
 + {\bar q}^{\mathord\uparrow}(x)
 - {\bar q}^{{\mathord\downarrow}}(x)]\>
dx,
\end{equation}
where $q^{\mathord\uparrow}(x)$ $\big[q^{\mathord\downarrow}(x)\big]$
is the quark distribution with flavour $q$ having spin (helicity)
parallel (anti parallel) to the nucleon spin.

It is well known that semileptonic decays of octet baryons can be
well described in the Cabibbo model, where one assumes that the
axial currents responsible for the semileptonic decays belong to an
$SU(3)$
octet. The diagonal matrix elements of these axial currents in this
model give the well-known connection to the Ellis-Jaffe Sum
Rule. They can be expressed as
\begin{equation}
\langle p,s|A_\mu^a|p,s\rangle  = 2 m_N s_\mu\cdot \Delta q_a
\label{bareax}
\end{equation}
with $s_\mu$ being the spin-vector ($s\cdot s=-1$, $p\cdot s=0$),
$A_\mu^a$ is an axial current defined by
$A_\mu^a=\bar q\gamma_\mu\gamma^5 T^a q$, with $T^a=\lambda^a/2$
being $SU(3)$ generators ($\lambda^a$ are the Gell-Mann matrices),
and $T^0$ being the identity matrix and
\begin{equation}
\begin{array}{rll}
2\Delta q_3 &= \Delta u-\Delta d           &=F+D   \\
2\sqrt 3\Delta q_8 &= \Delta u+\Delta d-2\Delta s &=3F-D  \\
\Delta q_0 &= \Delta u+\Delta d+\Delta s.  &       \\
\end{array}.
\label{currents}
\end{equation}
The axial coupling constants $F$ and $D$ can be fixed by fitting
to semileptonic decay data.

The non-singlet quantities $\Delta q_{3}$ and $\Delta q_{8}$ are
multiplied by a known QCD correction factors $(1 - \alpha_s / \pi)$
\cite{K179,K279} (the correction factor for the singlet quantity
$\Delta q_0$ is different) which for the sake of simplicity we shall
suppress throughout this paper.

Taking the EMC result (Eq.\ref{EMC}) at face value and assuming
$SU(3)$ symmetry(!), in combination with experimental results
for the neutron beta decay and the semileptonic decay data for
the octet baryons \cite{PDG92}, led to an unexpected result
\begin{equation}
\Delta q_{0} = 0.120 \pm 0.094(stat) \pm 0.138(syst),
\label{g0}
\end{equation}
which was consistent with zero. This can be interpreted such that
only a small fraction of the proton spin is carried by quarks.

Newer experimental data taken at CERN \cite{AAA94} and SLAC
\cite{AAB93} are to some degree controversial. The experimental data
for the polarized muon scattering on a deuterium target \cite{AAA94}
agrees with the earlier EMC data, with the Ellis-Jaffe integral
smaller than the prediction of the Ellis-Jaffe Sum Rule, the fraction
of the nucleon spin carried by quarks very small and appreciable
negative polarization of the strange quarks. On the other hand, the
result obtained in polarized electron scattering from the $^{3}He$
target \cite{AAB93} in the SLAC E142 experiment agrees with the
prediction of the Ellis-Jaffe Sum Rule, finding that a relatively
large fraction of the nucleon spin is carried by quarks and giving a
$\Delta s$ consistent with zero. The experimental results will be
summarized later when comparing with results obtained from our model.

In the naive $SU(6)$ quark model (NQM) $F=2/3$ and $D=1$, and as a
consequence $2\Delta q_{3} = 5/3$, $2\sqrt 3\Delta q_{8} = \Delta
q_{0} = 1$.  The 'experimental' result (Eq. \ref{g0}) is in marked
disagreement with the NQM. The NQM neglects the fact that the proton
constituents are highly relativistic \cite{CJJT74} and interacting
\cite{TW91} objects. Furthermore, not only quarks contribute to the
axial-vector singlet but also gluons. Unlike the flavour octet
currents, the singlet current $A_{\mu}^{0}$ has an anomalous
divergence (for a review see e.g. Ref.\cite{S191})
\begin{equation}
\partial^{\mu} A_{\mu}^{0} = \frac{N_{f} \alpha_{s}}{2\pi}
Tr {\cal G} {\tilde {\cal G}} ,
\end{equation}
where $N_{f}$ is the number of active flavours and ${\cal G}$ is the
gluonic field.  The triangle axial anomaly gives rise to an
independent gluonic contribution to the flavour singlet axial
current
\cite{AR88,CCM88,JM90,JPSSW90,EST90,S91,F91,GR91,EST91}.
As proposed by Efremov-Teryaev
\cite{ET88} and Altarelli-Ross \cite{AR88}  the parton model (PM)
values $\Delta u$, $\Delta d$ and $\Delta s$ should be modified
\begin{equation}
\Delta q(x,Q^2)\rightarrow \Delta q(x,Q^2)|_{PM} - \delta(x,Q^2) \; ,
\end{equation}
where \cite{BT93A}
\begin{equation}
\delta(x,Q^2) = {\alpha_{s} \over 2\pi}
\int_{x}^{1} C^{\Delta g}({x \over y},Q^2)\Delta g(y, Q^2) \; dy \; ,
\end{equation}
is a correction induced by the axial anomaly.  $C^{\Delta g}(z)$ is
the probability to find a quark of longitudinal momentum fraction $z$
in a gluon and appropriate polarization.  The total gluonic
contribution to $\Delta q$ is
\begin{equation}
\delta \equiv \int_{0}^{1} \delta (x, Q^2) =
{\alpha_{s}(Q^2) \over 2\pi} \;\Delta g(Q^2) \; .
\end{equation}
$\Delta g$ in the last formula is the fraction of the proton spin
carried by gluons. As a consequence $\Sigma_{PM}$ should be replaced
by $\Sigma_{PM} - 3 \delta$.  The coefficients $C_{\Delta g}$ have
been calculated \cite{ET88,AR88,CCM88,BINT91}.  On the other hand,
$\Delta g$ cannot be calculated and most of the authors tried to
adjust $\Delta g$ to describe experimental data (see for instance
\cite{GS95}).  There has been also much discussion over the
uniqueness of the separation into $\Delta q_{PM}$ and $\delta$ (see
for instance \cite{BINT91}). It has been argued \cite{JM90,B95} that
the anomaly induces also a non-perturbative gluonic contribution to
$\Delta q(x,Q^2)|_{PM}$.  Although it is very important in
understanding the Ellis-Jaffe Sum Rule violation, up to now no
definite treatment of the axial anomaly has been worked out.  We
shall come back to the problem of the anomaly when discussing the
mesonic effects.

\section{Semileptonic decays}

According to our present understanding, the weak semileptonic decays
of the octet baryons can be classified into two groups: either a
$d$-quark is transformed into a $u$-quark, or an $s$-quark is
transformed into a $u$-quark. The matrix elements of the
current operators 'responsible' for the semileptonic decays of the
baryons belonging to the octet can be parameterized in terms of
$q^2$-dependent form factors
\begin{equation}
\langle B_1|V_\mu+A_\mu|B_2\rangle=
\begin{array}[t]{rllll}
C\bar u_{B_1}\big[
       & f_1(q^2)\gamma_\mu
       & \displaystyle +i{f_2(q^2)\over m_1+m_2}\sigma_{\mu\nu}q^\nu
       & \displaystyle +{f_3(q^2)\over m_1+m_2} q_\mu & \\
  +    & g_1(q^2)\gamma_\mu\gamma^5
       & \displaystyle +i{g_2(q^2)\over m_1+m_2} \sigma_{\mu\nu}
q^\nu\gamma^5
       & \displaystyle +{g_3(q^2)\over m_1+m_2} q_\mu\gamma^5
       & \big] u_{B_2}. \\
\end{array}
\end{equation}
The factor $C$ here is the Cabibbo factor ($\sin \theta_C$ or
$\cos\theta_C$). At low momentum transfer only two terms, $f_{1}$
(vector) and $g_{1}$ (axial vector), are important. It is customary
to extract from experiments the ratio $g_{A}/g_{V} =
g_{1}(0)/f_{1}(0)$.

The semileptonic decays can be well described assuming the so-called
$SU(3)$ (Cabibbo) model\cite{R90}. Within this model the operators
for the $d \to u$ and $s \to u$ transitions can be expressed in
terms of $SU(3)$ group generators
\begin{eqnarray}
d\to u\qquad&& A^{1+i2}_\mu=\gamma_\mu\gamma^5
\left( T^1+iT^2 \right), \\
s\to u\qquad&& A^{4+i5}_\mu=\gamma_\mu\gamma^5
\left( T^4+iT^5 \right),
\end{eqnarray}
which are related to the familiar Gell-Mann matrices $T^k =
{\lambda^k/2}$.

Mesonic corrections lead to the renormalization of the axial-vector
coupling constants. The vector coupling constants are protected
against renormalization by vector current conservation. Mesonic
corrections to the axial-vector coupling constants have been taken
into account by calculating the loop corrections to the tree level
approximation according to Eq.(\ref{axial}). The corresponding
diagrams are shown in Fig.2. Preliminary results with the inclusion
of intermediate pseudoscalar mesons and associated octet and
decuplet baryons have been already presented elsewhere
\cite{Elba93}.

To perform numerical calculations within our model requires the
knowledge of the axial coupling constants for the bare octet and
decuplet baryons, vector mesons and transitions octet
$\leftrightarrow$ decuplet.  The transitions within the baryonic octet
are traditionally parameterized by the so-called anti-symmetric $F$
and symmetric $D$ coupling constants.  The axial coupling constant for
the transition within the decuplet ($H$) can be fixed by the relation
$2\langle\Delta^{++}|A^3_\mu|\Delta^{++}\rangle=H\cdot 2m_\Delta
s_\mu$. In analogy we define the coupling constant for the
interference diagram octet $\longleftrightarrow$ decuplet ($I$) as
$2\langle p|A^3_\mu|\Delta^0\rangle=2\langle\Delta^0|A^3_\mu|p\rangle=
I\cdot 2\sqrt{ m_N m_\Delta} s_\mu$.  The matrix elements of
axial-vector currents between pseudoscalar mesons vanish. They are,
however, finite for vector mesons.  Here the structure is analogous to
that of the baryonic octet.  We denote the corresponding constants as
$FV$ and $DV$. Due to parity-conservation the axial coupling constant
$FV$ vanishes. A special role is played here by the $\omega$ meson
which consists of an octet and a singlet part \cite{HHS89}. For
simplicity we neglect couplings to the singlet part, although they are
in principle present. The vector meson $\leftrightarrow$ pseudoscalar
meson interference terms have an octet structure analogous to the
other cases, with coupling constants called $FI$ and $DI$.

In the $SU(6)$ model, i.e. in the model in which all particles are
described by their $SU(6)$ wave functions \cite{C79}, the axial
coupling constants can easily be calculated
\begin{equation}
F={2\over3},\quad D=1,\quad H=1,\quad I=4{\sqrt{2}\over3},
\quad FV=0,\quad DV=1,\quad FI=1,\quad DI=0.
\label{su6coup}
\end{equation}

In Table 3 we present a list of all measured semileptonic decays of
the octet baryons. The experimental $g_A/g_V$ ratios are taken from
Refs.\cite{PDG92,F72}. In addition, we present the $SU(3)$ values for
$g_A$'s expressed in terms of the symmetric and asymmetric coupling
constants and $g_V$'s. Vector current conservation allows one to
extract the experimental $g_A$'s, which are presented in the last
column.  In Table 4 we present the result of our calculations for all
possible semileptonic decays. In the column named '$MC,SU(6)$' we show
the $g_A$'s calculated within our model with the $SU(6)$ axial-vector
coupling constants (Eq. \ref{su6coup}) for the bare hadrons. In the
column labeled '$MC,SU(3)$' similar results are shown with $F$ and $D$
fitted to the measured values of the axial-vector coupling
constants. For comparison, we show the results for the pure (no
mesonic corrections) $SU(6)$ model and pure $SU(3)$ model ($F$ and $D$
fitted to the experimental data from Table 3).  The $\chi^2$ values
presented in the last row for each model give an idea of the fit
quality. It is well known that the naive $SU(6)$ model gives a very
poor description of the experimental semileptonic decay data. On the
other hand, when fitting the $F$ and $D$ parameters an extremely good
description of the existing data can be achieved. It is commonly
believed that any correction to the $SU(3)$ model may only destroy the
nice agreement.  Inclusion of mesonic corrections with $SU(6)$ axial
coupling constants improves the description of the data dramatically
($\chi^2/N = 4369 \rightarrow \chi^2/N = 8.5$). An additional
variation of the $F$ and $D$ parameters improves the fit further. We
cannot allow for variation of the remaining parameters ($H$, $I$,
$DV$, $FI$), because a completely unrestricted fit could result in
unphysical values of parameters since the number of experimental data
points is limited to only 5.

In order to demonstrate the effect of different Fock components we
present in Table 5 the axial-vector coupling constants $g_A$
calculated with the inclusion of intermediate pseudoscalar mesons and
octet baryons $(oct,ps)$, with the additional inclusion of decuplet
baryons $(ps)$ and with the additional inclusion of vector mesons
$(all)$. In the case of the $SU(6)$ axial coupling constants
(Eq. \ref{su6coup}) the inclusion of pseudoscalar mesons--octet baryons
Fock components improves
the quality of the fit to the experimental $g_A$'s in
comparison to NQM (see the $\chi^2$ values in the last row of the
table) tremendously. The additional inclusion of intermediate decuplet
baryons
deteriorates the fit, increasing the $\chi^2$ value. In this context
it is worth mentioning that the contribution of the octet-decuplet
interference terms to the axial-vector coupling constant for neutron
beta-decay $g_A^{n\to p}$ obtained in our calculation is much smaller
than in the Cloudy Bag Model\cite{ST88}. Adding vector mesons again
improves the fit to the experimental axial-vector coupling constants.

Summarizing, taking into account both pseudoscalar and vector meson
corrections does not lead to any significant deviations from the
experimental data for semileptonic decays. The quality of the fit is
comparable to that of the original Cabibbo model. Therefore, the
experimental data for semileptonic decays does not contradict the
meson cloud model of the nucleon, which one might naively expect. We
do not predict any essential difference between the $SU(3)$ Cabibbo
model and our model for any so far unmeasured transitions.

The meson radiative corrections were recently calculated in the
framework of baryon chiral perturbation theory including pseudoscalar
mesons with both intermediate baryon octet \cite{JM911} and
decuplet \cite{JM912} components. In order to simplify the
calculation the baryon fields were treated there as heavy static
fermions. Although the details of the calculation in
Refs.\cite{JM911,JM912} differ from ours, similar conclusions have
been drawn. In comparison to Refs.\cite{JM911,JM912}, where the
experimental error bars where increased in calculating their $\chi^2$
values, our agreement with the data is much better. If we increase
the error bars as in Refs.\cite{JM911,JM912} we would get $\chi^2/N$
of the order of 0.1 - 0.4, compared to $\sim 2$ in
Refs.\cite{JM911,JM912}.

\section{Meson Cloud Effects on the Spin Structure of the Nucleon}

In section II we discussed how the matrix elements of axial currents
of the dressed nucleon are related to the corresponding matrix
elements of the constituents (intermediate baryons and mesons) in the
convolution approach.  Eq.(\ref{currents}) allows one to relate axial
current matrix elements to quark polarizations.  Then the Ellis-Jaffe
Sum Rule can be expressed in terms of axial-vector matrix elements
$\Delta q_0$, $\Delta q_3$ and $\Delta q_8$ (Eq. \ref{currents})
\begin{equation}
S_{EJ}^p={1\over9}\Delta q_0+{1\over6}\Delta
q_3+{1\over6\sqrt3}\Delta q_8.
\end{equation}
In the same way the Bjorken Sum Rule becomes
\begin{equation}
S_{B} = \int_0^1 [ g_1^p(x) -g_1^n(x) ] dx = {1\over3}\Delta q_3
\end{equation}
which can be expressed in terms of the axial-vector coupling constant
for neutron beta decay
\begin{equation}
S_B={1\over 6}g_A^{n\to p}=0.210.
\label{bjorken}
\end{equation}

Practical calculation of axial-vector matrix elements (see
Eq. \ref{axial}) requires additional assumptions about the axial
properties of the bare particles. We will assume
$\Delta q_0=2\sqrt3\Delta q_8$ for all non-strange bare particles,
which is an extension of the original Ellis-Jaffe Ansatz $\Delta s=0$
\cite{EJ74}. In analogy to the previous section we will consider two
models: the $SU(6)$ model, with axial coupling constants given by
Eq.(\ref{su6coup}) and the $SU(3)$ model with $F$ and $D$ fitted to
the semileptonic data.

In Table 6 we present our results for the Ellis-Jaffe Sum Rule for
the proton $S_{EJ}^p$, the Ellis-Jaffe Sum Rule for the neutron
$S_{EJ}^n$, the Bjorken Sum Rule $S_B$ and $\Delta q_0$ in both
models with different Fock states included:
(a) tree level, (b) octet baryons and pseudoscalar
mesons $(oct,ps)$, (c) octet and decuplet baryons and pseudoscalar
mesons ($ps$) and (d) the same with an additional inclusion of
vector mesons ($all$).

In the case of the $SU(6)$ model inclusion of the $(oct,ps)$ Fock
states brings the bare $SU(6)$ value of $5/18 = 0.278$ for the proton
Ellis-Jaffe Sum Rule almost half way down to the experimental result
\cite{A89}. Inclusion of the intermediate decuplet baryons enhances
$S_{EJ}^p$ a little bit, and conversely the addition of vector mesons
decreases $S_{EJ}^p$ slightly. Even with the inclusion of all Fock
states we are not able to describe the experimental EMC result for
$S_{EJ}^p$.

It would be, however, too naive to expect that our model alone can
account for all spin problems; other effects, like the famous axial
anomaly (see section V), should be present as well.  Currently it
is, however, very difficult to give a quantitative estimate of the
axial anomaly effect.  On the purely phenomenological side one could
try to map the $x$ dependence of the axial anomaly contribution to
$g_1$ by ascribing all the deficiencies of the existing models with
respect to the experimental data to the axial anomaly.  Then one can
write
\begin{equation}
g_{1}^{p} = g_{1}^{p}(x)|_{model} + \Delta g_{1}(x) \; , \qquad
g_{1}^{n} = g_{1}^{n}(x)|_{model} + \Delta g_{1}(x) \; ,
\end{equation}
where $\Delta g_{1}(x)$ is the contribution of the axial anomaly.
In Ref.\cite{BT93B} the unknown contribution was ascribed
to the deficiency of the MIT bag model. Such an estimate of
the axial anomaly must rely on the model calculation of
$g_{1}^{p}(x)|_{model}$ and $g_{1}^{n}(x)|_{model}$.  In the present
paper we have concentrated on the role of mesonic corrections,
therefore considered only very simple models of the bare nucleons
(baryons). It has been shown very recently \cite{SHT95} that by
combining the mesonic effects calculated according to the present
paper together with one version of the Adelaide group bag model
\cite{ST94} an impressive agreement with the EMC \cite{A289} and the
SLAC \cite{A143} data can be obtained at $x > 0.1$.  The meson cloud
scenario leaves therefore much less phenomenological room for the
axial anomaly contribution, at least at $x > 0.1$. There is still
some room left at $x < 0.1$, i.e.  precisely in the region where the
gluonic contribution is expected to play an important role
\cite{BT93A}.

The Bjorken Sum Rule is independent of the anomaly.  Here we get a
good agreement with the classical value Eq.(\ref{bjorken}) (without
QCD corrections) when including only intermediate octet baryons with
pseudoscalar mesons as well as in the full model. In this context the
similarity to the Gottfried Sum Rule is worth noting. In order to
compare $S_B$ with the experimental results of the SLAC and SMC
experiments (see lower panel of Table 6), higher order perturbative
corrections have to be included \cite{EK95}. Perturbative QCD
corrections to the Bjorken Sum Rule have been calculated up to
$O((\alpha_s/\pi)^3)$ \cite{LV91}
\begin{equation}
S_B={1\over6} g_A\left[1-{\alpha_s(Q^2)\over\pi}
  -3.5833\left({\alpha_s(Q^2)\over\pi}\right)^2
  -20.2153\left({\alpha_s(Q^2)\over\pi}\right)^3\right].
\end{equation}
In the case of the SLAC experiment also higher twist effects probably
play an important role \cite{EK95}.  Once higher-order perturbative
corrections are taken into account we get a good agreement with the
result of the SMC analysis which include all available proton
and deuteron data \cite{AAA94}.

The classical $SU(6)$ model is attractive due to its
simplicity.  In practice, in order to describe different data it is
necessary to break this symmetry and allow $F$ and $D$ to be
fitted to the semileptonic decay data.
We also follow this line here.  Then we get $S_{EJ}^p$ closer
to the experimental EMC result with the 'best' result when only
intermediate octet baryons and pseudoscalar mesons are included. In
this case $S_{EJ}^n$ becomes negative, which seems to be the case
experimentally \cite{AAA93}. Of course the Bjorken Sum Rule is
fulfilled here in all cases, which is due to the relation between
$S_B$ and neutron beta decay constant $g_{A}^{n \rightarrow p}$
(Eq. \ref{bjorken}).

Also the calculated Ellis-Jaffe Sum
Rule has to be corrected for perturbative QCD effects. It has been
shown \cite{K80} that the singlet ($\Gamma^S$) and the non-singlet
($\Gamma^{NS}$) contributions to the integral of the Ellis-Jaffe Sum
Rule get different corrections
\begin{equation}
\Gamma^{S} = \Gamma^{S}_{0}
\left( 1 - {33 - 8N_{f} \over 33 - 2N_{f}}
 {\alpha_{s} \over \pi} \right) \, ,
\qquad
\Gamma^{NS} = \Gamma^{NS}_{0}
\left(1 -  {\alpha_{s} \over \pi}\right) \, .
\end{equation}
For example for $Q^2 = 2\,GeV^2$ $S_{EJ}^{p}$ calculated within
our model (see Table 6) is reduced:
$0.220 \rightarrow 0.201$ $(SU(6), all)$ and
$0.179 \rightarrow 0.163$ $(SU(3), all)$.
Here we have taken $\alpha_{s}(2\,GeV^2) = 0.371$ \cite{EK93}.
Although substantial, the pQCD corrections does not allow to bring
our results down to the experimental $S_{EJ}^{p}$.

In all models considered, the singlet axial matrix element $\Delta
q_{0}$ and $\Delta s$ are far from the one obtained from the
EMC analysis(!) \cite{A89}. We have stressed the word
analysis, as the EMC result is biased by the assumption of the
$SU(3)$ symmetry.  Since our model violates $SU(3)$ symmetry, $\Delta
q_0$ does not need to coincide with the value extracted by the
EMC. In our model $\Delta q_0$ is reduced mainly due to the $\pi N$
Fock state admixture, which causes a partial depolarization of the
spin as discussed in Refs.\cite{Z93}. The depolarization is
caused by a cancellation of the spin-preserving and spin-flipping
contributions to the $\pi N$ Fock state.

Please note that our model predicts $\Delta s > 0$ in contrast to the
negative value found from the EMC analysis. In the constituent quark
models the nucleon does not contain any internal strangeness and as a
consequence $g_{NN\phi} =0$ because $\phi$ is known to be an almost
pure $s \bar s$ state. This is also the case of the so-called $SU(6)$
related coupling constants \cite{HHS89} which we have used throughout
this paper.  However, $\phi$ could couple to the nucleon through
$K\bar K$ very much the same as the $\rho$ meson couples to the
nucleon via $\pi\pi$ \cite{HP75}.  Due to its structure the $\phi$
meson could be a good candidate to understand the problem of the
``missing'' polarized strangeness.  To estimate a maximal possible
effect of $\phi$ we have taken $g_{NN\phi}$ from the vector meson
dominance model \cite{IKL84}.  Then we get extra $\Delta s \approx
0.005 > 0$, which is rather small and of opposite sign to the
"experimental" $\Delta s$.

\section{Conclusions}

In the light of recent experiments on the deep-inelastic lepton
scattering by nucleons, the understanding of the nucleon structure
has become a hot topic in particle and nuclear physics. The recent
observation of Gottfried Sum Rule breaking strongly suggests a
flavour asymmetry of light sea quarks in the nucleon. The asymmetry
occurs in a natural way within the meson cloud model.

The meson cloud model, as in any phenomenological model, requires
specification of some external parameters which cannot be calculated
at the present time from a more microscopic theory. Analysis of the
model over the past few years has shown that the results are rather
sensitive to these parameters. At the present stage, it seems to be
impossible to derive these parameters from the underlying QCD. In the
present paper we have tried to set limits on these parameters by
applying the model to high-energy baryon production data. The
application of the MCM to these reactions is consistent with its
application to the nucleon structure.  In
practical calculations one has to make an ansatz for the vertex form
factors. It was argued \cite{Z92} that one should use form factors
which guarantee certain symmetries of the longitudinal momentum
distribution functions of virtual mesons and baryons. In the present
work we followed this approach. In contrast to form factors often
used in traditional nuclear physics calculations, the 'symmetric form
factors' give a good description of the experimental data. We find an
universal cut-off parameter for all Fock states involving octet
baryons.  This leads to important differences in comparison to other
calculations, where universality has been assumed for the
$t$-dependent form factors.

After we had fixed the parameters of the vertex form factors in the
hadronic sector, we have looked at the consequences in unpolarized and
polarized deep-inelastic scattering and for the semileptonic
decays. We find a few interesting results.  The value of the Gottfried
Sum Rule obtained from our model ($S_G=0.224$) is in good agreement
with the experimental result obtained by NMC ($S_G=0.24\pm
0.016$). The meson cloud model predicts a $\overline{u} -
\overline{d}$ asymmetry concentrated at rather small $x$ in impressive
agreement with a recent global fit to the world data on the Drell-Yan
and DIS \cite{MSR93}.

Since in our model the probability of the strange meson---strange
baryon Fock components is substantially reduced in comparison to the
non-strange counterparts, we obtain a significant reduction of the
strange sea quark distributions. We find $s(x)\ne \bar s(x)$ in
contrast to the customary assumption $s(x)=\bar s(x)$. The effect
found is small and differs from naive expectations.

The mesonic corrections lead to renormalization of axial-vector
current matrix elements. Large one-loop corrections, which
explicitly violate $SU(3)$ symmetry, are in surprisingly good
agreement with semileptonic data even with $SU(6)$ axial coupling
constants for bare particles.

We find that a significant fraction (20\%) of the nucleon spin is
carried by the angular momentum of the mesonic cloud.  Although the
meson cloud model does not reproduce completely the EMC result for the
Ellis-Jaffe Sum Rule, the contribution of the meson cloud cannot be
neglected in the total balance of the proton spin, especially in the
context of the success of the meson cloud in the explanation of the
Gottfried Sum Rule violation and $\bar u(x)-\bar d(x)$ asymmetry. The
presence of higher Fock components seems, however, not sufficient to
resolve the spin crisis. Here, probably other effects, to mention only
the axial anomaly, play an important role.

Summarizing, in our view the meson cloud model has many attractive
features and can account for the description of many experimental
data.  Here, we have discussed only some aspects. It should be
mentioned that this picture of the nucleon provides, in addition, a
good description of nucleon electric polarizabilities \cite{BKM92}
and that it has a close connection to very successful models of
low-energy hadron-hadron scattering \cite{MHE87,HHS89}.

Finally we want to point out that the same model (with the same set of
parameters which has been determined from quite different phenomena)
turned out to be very successful \cite{HNSS94} in the description of
the new NA51 CERN data for the dilepton production in the
proton-proton and proton-deuteron collisions \cite{B94}.  The model
provides also a good description of the production of slow protons on
the ``neutron'' (extracted from the deuteron target \cite{BDT94}) in
the charge-current neutrino and antineutrino DIS reactions
\cite{SBD94}. The model discussed here gives also unique predictions
for the semi-inclusive DIS $e+p \rightarrow e'+n+X$ which will be
tested soon at HERA \cite{HLNSS94}. Such a measurement is being
prepared by the ZEUS Collaboration which is installing the forward
neutron calorimeter \cite{LF92,ZEUS94}. A pilot experiment on
detecting neutrons from the beam-gas interactions has already been
performed and the result is in good agreement with the pion-exchange
predictions \cite{ZEUS94}. Like any model, the meson cloud model
requires further testing. In this respect the planned Fermilab
experiment measuring the relative dilepton yield in proton-proton and
proton-deuteron interactions \cite{G92,SSG94} seems to be very
important and promising.

\bigskip
{\bf Acknowledgments}

Very useful discussion with S.D. Bass, N.N. Nikolaev, A.W. Thomas and
V.R. Zoller is gratefully acknowledged. We wish to thank D. von
Harrach for drawing our attention to the problem of the $s$-$\bar s$
asymmetry. We are indebted to B. Gibson and Ch. Goodman for carefully
reading
the manuscript and very useful discussion. This work was supported in
part by the Polish KBN grant 2 2409 9102.

\begin{appendix}
\section{Lagrangians}
Here we present the interaction langragians we employ in our
calculations.  They are usually used in meson exchange models
\cite{MHE87}.  $\phi$ denotes a spin-$1/2$ field ($N$), $\psi$ a
spin-$3/2$ field ($\Delta$) of Rarita-Schwinger form; $\pi$
denotes a pseudoscalar field and $\theta$ a vector field
($\rho,\omega$):

\begin{eqnarray}
   {\cal L}_1&=&g\cdot i\bar\phi{\gamma_5}\pi\phi, \\ {\cal
L}_2&=&f\cdot\bar\phi\partial_\mu\pi\psi^\mu+h.c.\>,\\ {\cal
L}_3&=&g\cdot\bar\phi\gamma_\mu\theta^\mu\phi
+f\cdot\bar\phi\sigma_{\mu\nu}\phi
(\partial^\mu\theta^\nu-\partial^\nu\theta^\mu), \\ {\cal L}_4&=&
f\cdot i\bar\phi{\gamma_5}\gamma_\mu\psi_\nu
(\partial^\mu\theta^\nu-\partial^\nu\theta^\mu)+h.c.\>.
\end{eqnarray}
The anti-symmetric tensor $\sigma_{\mu\nu}$ here is defined as
$\sigma_{\mu\nu}={i\over 2}[\gamma_\mu,\gamma_\nu]$.

\section{Vertex functions}

Here we collect our results for the helicity dependent vertex
functions $V_{IMF}^{\lambda\lambda'}(y,k^2_\perp)$. $y$ here denotes
the longitudinal momentum fraction of the baryon in the nucleon;
${\vec k_\perp}=(k_\perp\cos\varphi,k_\perp\sin\varphi)$ the
transverse momentum of the baryon with respect to the nucleon
momentum. The contributions are listed according to particle
helicities $(1/2\to \lambda,\lambda')$, with $\lambda$ and $\lambda'$
being the baryon and meson helicities, respectively.

\bigskip
\noindent a.) Transitions for ${\cal L}_1$
($N\pi$, $N\eta$, $\Sigma K$, $\Lambda K$)
\medskip
\begin{tabbing}
$\displaystyle+{1\over2}\;\;$ \= $\displaystyle0\;\;\quad$ \=
  $\displaystyle{g\over2}
    {ym_N-m_B\over{\sqrt{\vphantom{I} y m_Nm_B}}}$ \\[2mm]

$\displaystyle-{1\over2}$     \> $\displaystyle0$          \>
  $\displaystyle{g e^{-i\varphi}\over2}
   {k_\perp\over{\sqrt{\vphantom{I} y m_Nm_B}}}$ \\[2mm]
\end{tabbing}
\medskip

%\pagebreak
\noindent b.) Transitions for ${\cal L}_2$
($\Delta\pi$, $\Sigma^* K$)

\medskip
\begin{tabbing}
$\displaystyle+{3\over2}\;\;$ \= $\displaystyle0\;\;\quad$ \=
  $\displaystyle-{f e^{+i\varphi}\over2\sqrt2}{k_\perp(ym_N+m_B)
      \over y{\sqrt{\vphantom{I} y m_Nm_B}}}$ \\[2mm]

$\displaystyle+{1\over2}$\>$\displaystyle0$\>
  $\displaystyle {f\over2\sqrt6}
      {(ym_N+m_B)^2(ym_N-m_B)+k^2_\perp(ym_N+2m_B)\over
       ym_B{\sqrt{\vphantom{I} y m_Nm_B}}}$ \\[2mm]

$\displaystyle-{1\over2}$\>$\displaystyle0$\>
  $\displaystyle {f e^{-i\varphi}\over2\sqrt6}
      {k_\perp[(ym_N+m_B)^2-3m_B(ym_N+m_B)+k^2_\perp]\over
       ym_B{\sqrt{\vphantom{I} y m_Nm_B}}}$ \\[2mm]

$\displaystyle-{3\over2}$\>$\displaystyle0$\>
  $\displaystyle -{f e^{-2i\varphi}\over 2\sqrt2}
    {k^2_\perp\over y{\sqrt{\vphantom{I} y m_Nm_B}}}$ \\[2mm]
\end{tabbing}
\medskip

\noindent c.) Transitions for ${\cal L}_3$
($N\rho$, $N\omega$, $\Sigma K^*$, $\Lambda K^*$)

\medskip
\begin{tabbing}
$\displaystyle+{1\over2}\;\;$\=$\displaystyle+1\quad$\=
  $\displaystyle {g e^{+i\varphi}\over\sqrt2}
   {k_\perp\over(1-y){\sqrt{\vphantom{I} y m_Nm_B}}}
      -f\sqrt2 e^{+i\varphi}
        {k_\perp m_N\over{\sqrt{\vphantom{I} y m_Nm_B}}}$ \\[2mm]

$\displaystyle+{1\over2}$\>$\displaystyle0$\>
  $\displaystyle {g\over2}{k^2_\perp+m_Nm_B(1-y)^2-ym_M^2
              \over(1-y)m_M{\sqrt{\vphantom{I} y m_Nm_B}}}$ \\[2mm]

\>\> $\displaystyle-{f\over2}{(ym_N-m_B)(y^2m_N^2-y(m_N^2+m_B^2+m_M^2)
                  +m_B^2+k^2_\perp)
                   \over ym_M{\sqrt{\vphantom{I} y m_Nm_B}}}$ \\[2mm]

$\displaystyle+{1\over2}$\>$\displaystyle-1$\>
  $\displaystyle {g e^{-i\varphi}\over\sqrt2}
               {yk_\perp\over(1-y){\sqrt{\vphantom{I} y m_Nm_B}}}
    +f\sqrt2 e^{-i\varphi}{k_\perp m_B\over
       {\sqrt{\vphantom{I} y m_Nm_B}}}$ \\[2mm]

$\displaystyle-{1\over2}$\>$\displaystyle+1$\>
  $\displaystyle -{g\over\sqrt2}
         {ym_N-m_B\over{\sqrt{\vphantom{I} y m_Nm_B}}}$ \\[2mm]

\>\> $\displaystyle   -f\sqrt2{k^2_\perp-(m_N+m_B)(1-y)(ym_N-m_B)
           \over(1-y){\sqrt{\vphantom{I} y m_Nm_B}}}$ \\[2mm]

$\displaystyle-{1\over2}$\>$\displaystyle0$\>
  $\displaystyle -{g e^{-i\varphi}\over2}{k_\perp(m_N-m_B)\over
       m_M{\sqrt{\vphantom{I} y m_Nm_B}}}$ \\[2mm]

\>\> $\displaystyle -{f e^{-i\varphi}\over2}{k_\perp(1+y)
           (y^2m_N^2-y(m_N^2+m_B^2+m_M^2)+m_B^2+k^2_\perp)
            \over y(1-y)m_M{\sqrt{\vphantom{I} y m_Nm_B}}}$ \\[2mm]

$\displaystyle-{1\over2}$\>$\displaystyle-1$\>
  $\displaystyle f\sqrt2 e^{-2i\varphi}{-2}{k^2_\perp
     \over(1-y){\sqrt{\vphantom{I} y m_Nm_B}}}$ \\[2mm]
\end{tabbing}
\medskip

\noindent d.) Transitions for ${\cal L}_4$
($\Delta\rho$, $\Sigma^* K^*$)

\medskip
\begin{tabbing}
$\displaystyle+{3\over2}\;\;$\=$\displaystyle+1\quad$ \=
  $\displaystyle -{f e^{+2i\varphi}\over2}{k^2_\perp\over y(1-y)
      {\sqrt{\vphantom{I} y m_Nm_B}}}$ \\[2mm]

$\displaystyle+{3\over2}$\>$\displaystyle0$\>
  $\displaystyle {f e^{+i\varphi}\over\sqrt2}{k_\perp m_M\over
      (1-y){\sqrt{\vphantom{I} y m_Nm_B}}}$ \\[2mm]

$\displaystyle+{3\over2}$\>$\displaystyle-1$\>
  $\displaystyle {f\over2}{m_Nm_B(1-y)^2-ym_M^2\over
     (1-y){\sqrt{\vphantom{I} y m_Nm_B}}}$ \\[2mm]

$\displaystyle+{1\over2}$\>$\displaystyle+1$\>
  $\displaystyle {f e^{+i\varphi}\over2\sqrt3}
         {k_\perp[k^2_\perp-2(1-y)m_B^2]
           \over y(1-y)m_B{\sqrt{\vphantom{I} y m_Nm_B}}}$ \\[2mm]

$\displaystyle+{1\over2}$\>$\displaystyle0$\>
  $\displaystyle -{f\over\sqrt6}{m_M[k^2_\perp+m_B(1-y)(ym_N-m_B)]
               \over(1-y)m_B{\sqrt{\vphantom{I} y m_Nm_B}}}$ \\[2mm]

$\displaystyle+{1\over2}$\>$\displaystyle-1$\>
  $\displaystyle {f e^{-i\varphi}\over2\sqrt3}
          {k_\perp[ ym_M^2-2m_Nm_B(1-y)]
               \over (1-y)m_B{\sqrt{\vphantom{I} y m_Nm_B}}}$ \\[2mm]

$\displaystyle-{1\over2}$\>$\displaystyle+1$\>
  $\displaystyle {f\over2\sqrt3}
          {2(1-y)m_Bk^2_\perp+m_Nm_M^2 y^3-(1-y)^2
      m_B^3 \over y(1-y)m_B{\sqrt{\vphantom{I} y m_Nm_B}}}$ \\[2mm]

$\displaystyle-{1\over2}$\>$\displaystyle0$\>
  $\displaystyle {f e^{-i\varphi}\over\sqrt6}
    {k_\perp m_M(ym_N-(1-y)m_B)
             \over (1-y)m_B{\sqrt{\vphantom{I} y m_Nm_B}}}$ \\[2mm]

$\displaystyle-{1\over2}$\>$\displaystyle-1$\>
  $\displaystyle {f e^{-2i\varphi}\over2\sqrt3}
  {k^2_\perp m_N\over(1-y)m_B{\sqrt{\vphantom{I} y m_Nm_B}}}$ \\[2mm]

$\displaystyle-{3\over2}$\>$\displaystyle+1$\>
  $\displaystyle {f e^{-i\varphi}\over2}{k_\perp m_B(1-y)\over
       y{\sqrt{\vphantom{I} y m_Nm_B}}}$ \\[2mm]

$\displaystyle-{3\over2}$\>$\displaystyle0$\> $0$\\[2mm]

$\displaystyle-{3\over2}$\>$\displaystyle-1$\>$0$\\[2mm]
\end{tabbing}
\end{appendix}

%\bibliographystyle{my}
%\bibliography{dis}

\begin{figure}
\caption{ Deep inelastic scattering on a virtual meson (a) or
baryon (b).}
\end{figure}

\begin{figure}

\caption{ Corrections to an axial current in the MCM.}
\end{figure}

\begin{figure}
\caption{ One-Boson-Exchange diagrams for $n$ and $\Lambda$
production.}
\end{figure}

\begin{figure}
\caption{ Differential cross section for $pp\to nX$ (Fig.4a)
and $pp\to\Lambda X$ production \protect\cite{FM76,B178}(Fig.4b).
Shown are
the OBE contributions: pseudoscalar mesons (dashed), vector mesons
(dotted) and their sum (solid).}
\end{figure}

\begin{figure}
\caption{  The same as in Fig.4a for an traditional dipole
form factor with $\Lambda=1.2\>GeV$.}
\end{figure}

\begin{figure}
\caption{One-Boson-Exchange diagram for $\Delta^{++}$
production.}
\end{figure}

\begin{figure}
\caption{ Differential cross section for $pp\to\Delta^{++} X$
production.  The experimental spectra shown are from
Ref.~\protect\cite{B179}.
The dashed line corresponds to $\pi$-exchange, the dotted line to
$\rho$-exchange and the solid line to the sum of both
contributions. The two different scales correspond to different
background subtractions. The calculated results are shown with
respect to the right scale.}
\end{figure}

\begin{figure}
\caption{ The $y$-integrated $pp\to\Delta^{++}$ and $\bar
pp\to\Delta^{++}$ cross sections \protect\cite{B179,B180}:
$\pi$-exchange (dashed), $\rho$-exchange (dotted) and their sum
(solid).}
\end{figure}

\begin{figure}
\caption{The result of the fit of the $R_{sea}$ factor to the CCFR
experimental data for $\bar q(x)$.}
\end{figure}

\begin{figure}
\caption{ (a) The $x(\bar d - \bar u)$ asymmetry as a function of the
Bjorken variable $x$ calculated within our model with three different
pion structure functions: NA3 LO \protect\cite{B83} (dotted), NLO
\protect\cite{SMRS92} (solid) and of the GRV model
\protect\cite{GRV92_pi} (dashed).\hfil\break (b) The same for recent
experimental parameterizations with explicit inclusion of the flavour
asymmetry: MSR($D_0'$) \protect\cite{MSR93} (dotted), MSR($D_{-}'$)
\protect\cite{MSR93} (dash-dotted) and the newest parameterization of
the same group MSR(A) \protect\cite{MSR94} (dashed). The full line
(MCM) is again a prediction of our model.}
\end{figure}

\begin{figure}
\caption{ The difference $x({1\over2}(\bar u+\bar d)-\bar s)$ as
function of $x$.  The points with error bars have been extracted by
Kumano using results from the E615 collaboration \protect\cite{K191}.
The dashed line
is the MCM result obtained with valence quarks only. Inclusion of the
sea-quarks in mesons and baryons (as described in the text) gives the
solid line.}
\end{figure}

\begin{figure}
\caption{ The contributions of $\bar s_{val}^M$ and $s_{val}^B$ to
the nucleon $\bar s(x)$ and $s(x)$ quark distributions, respectively,
for LO parameterizations \protect\cite{B83},\protect\cite{O91}
(dashed) and NLO parameterizations\protect\cite{SMRS92},
\protect\cite{MSR93} (solid).\hfil\break (b) $s(x) - \bar s(x)$ for
two sets of structure functions: $\pi$ \protect\cite{B83} + N
\protect\cite{O91} (dashed) and $\pi$ \protect\cite{SMRS92} + N MSR
$S_{0}'$ \protect\cite{MSR93} (solid). \hfil\break (c) The
contributions from $s_{sea}^M$, $s_{sea}^B$ and $s_{sea}^{N'}$ (bare
nucleon) to the strange sea of the nucleon.  Details are described in
the text. The structure functions taken from
Refs.\protect\cite{MSR93},\protect\cite{SMRS92}.  For comparison we
show the ``experimental'' result of the CCFR collaboration (dashed
line).}
\end{figure}

\begin{table}
\begin{tabular}{lddd}
                      & exponential & monopole & dipole \\
\hline
$\Lambda_\pi \>(\>GeV)$& 1.08 & 1.9      & 2.2    \\
$\Lambda_\rho\>(\>GeV)$& 1.08 & 1.4      & 2.0    \\
\hline
$f_{p\pi^0/p}$        & 0.204  & 0.210   & 0.219 \\
$\Delta f_{p\pi^0/p}$ & 0.017  & 0.015   & 0.014 \\
$f_{p\rho^0/p}$       & 0.062  & 0.075   & 0.060  \\
$\Delta f_{p\rho^0/p}$&$-$0.038&$-$0.042&$-$0.034  \\
\end{tabular}
\caption{
Cut-off parameters for the different functional forms of
form factors obtained by fitting to the neutron production data and
the corresponding first moments of splitting functions.}
\end{table}

\begin{table}
\begin{tabular}{lddd}
model       & $Z$   & $A_{\bar u-\bar d}$&$S_G$ \\
\hline
$tree  $    & 1.    & 0.       & 0.333 \\
$oct,ps$    & 0.755 & $-$0.155 & 0.230 \\
$ps   $     & 0.697 & $-$0.119 & 0.254 \\
$all  $     & 0.580 & $-$0.163 & 0.224 \\
\end{tabular}
\caption{ The Gottfried Sum Rule and the $\bar u$-$\bar d$ asymmetry
in different models. Model $(oct,ps)$ includes all Fock states with
octet baryons and pseudoscalar mesons, $(ps)$ all Fock states with
octet and decuplet baryons and pseudoscalar mesons and $(all)$ all
discussed Fock states. The experimental value of the Gottfried Sum
Rule is $0.24\pm 0.016$ \protect\cite{A91}.  }
\end{table}

\begin{table}
\begin{tabular}{lcrcccc}
\multicolumn{3}{c}{decay} &
 $g_A\ SU(3)$ & $g_V\ SU(3)$ & $g_A/g_V\ exp$
& $g_A\ exp$ \\
\hline
$n$        &$\to$& $p$       & $F+D$ & 1 &
   $1.2573\pm0.0028$ & $1.273\pm0.0028$ \\
$\Sigma^-$ &$\to$& $\Lambda$ & $2D/\sqrt6$ & 0 &
   & $0.60\pm0.03$ \\
$\Lambda$ &$\to$& $p$ & $-(3F+D)/\sqrt6$ & $-\sqrt{3/2}$ &
   $ 0.718\pm0.015$ & $-0.857\pm0.018$ \\
$\Xi^-$ &$\to$& $\Lambda$ & $(3F+D)/\sqrt6$ & $\sqrt{3/2}$ &
   $ 0.25\pm0.05$ & $0.31\pm0.06$ \\
$\Sigma^-$ &$\to$& $n$ & $D-F$ & $-1$ &
   $-0.34\pm0.05$ & $0.34\pm0.05$ \\
\end{tabular}
\caption{List of all measured semileptonic decays of baryons and the
corresponding coupling constants in the $SU(3)$ model. The data were
taken from \protect\cite{PDG92} and \protect\cite{F72}
(for $\Sigma^-\to\Lambda$).}
\end{table}

\begin{table}
\begin{tabular}{lcrddddl}
\multicolumn{3}{c}{decay}
& $SU(6)$ & $SU(3)$ & $MC,SU(6)$ & $MC,SU(3)$ & $g_A$ $exp$ \\
\hline
$n$&$\to$&$p$
& 1.67    & 1.257   &  1.241     &   1.257    & $1.2573\pm0.0028$ \\
$\Sigma^+$&$\to$&$\Lambda$
& 0.82    &  0.67   &   0.66     &    0.74    &      \\
$\Sigma^-$&$\to$&$\Sigma^0$
& 0.94    &  0.62   &  0.77      &    0.64    &      \\
$\Sigma^-$&$\to$&$\Lambda$
& 0.82    &  0.67   &  0.65      &    0.75    & $0.60\pm0.03$ \\
$\Sigma^+$&$\to$&$\Sigma^0$
& $-$0.94 & $-$0.62 & $-$0.77    &  $-$0.64   &      \\
$\Xi^-$&$\to$&$\Xi^0$
& 0.33    &  0.38   &  0.27      &    0.49    &      \\
\hline
$\Lambda$&$\to$&$p$
& $-$1.22 &$-$0.87  & $-$0.96    &  $-$0.89   & $-0.857\pm0.018$ \\
$\Sigma^0$&$\to$&$p$
& 0.24    &  0.27   &   0.19     &     0.31   &      \\
$\Sigma^-$&$\to$&$n$
& 0.33    &  0.38   &   0.27     &     0.49   & $0.34\pm0.05$ \\
$\Xi^0$&$\to$&$\Sigma^+$
& 1.67    &  1.26   &   1.37     &     1.39   &      \\
$\Xi^-$&$\to$&$\Lambda$
& 0.41    &  0.20   &  0.35      &    0.16    & $0.31\pm0.06$ \\
$\Xi^-$&$\to$&$\Sigma^0$
& 1.18    &  0.89   &   0.97     &    0.98    &      \\
\hline
\multicolumn{3}{c}{$\chi^2/N$}
& 4369    &  2.0    &  8.5      &    6.5     &      \\
\end{tabular}
\caption{ A list of all possible semileptonic decays of baryons
within the nucleon octet. The axial couplings $g_A$ has been
calculated at the tree-level in the $SU(6)$ model ($F=2/3$ and $D=1$)
and in the $SU(3)$ model with $F$ and $D$ ($F=0.44$, $D=0.82$)
fitted.  Also shown are the results with explicit inclusion of the
mesonic corrections with $SU(6)$ and $SU(3)$ axial-vector constants
for the bare baryons. In the $SU(3)$ case we find $F=0.53$ and
$D=1.15$.}
\end{table}

\begin{table}
\begin{tabular}{rcldddddd}
 &&&\multicolumn{3}{c}{$SU(6)$} & \multicolumn{3}{c}{$SU(3)$} \\
\cline{4-9}
\multicolumn{3}{c}{decay}
& $oct,ps$ & $ps$ & $all$ & $oct,ps$ & $ps$ & $all$ \\
\hline
$n$&$\to$&$p$
& 1.247  & 1.454  & 1.241 &   1.257& 1.257 &  1.257 \\
$\Sigma^-$&$\to$&$\Lambda$
&   0.70 &   0.77 &   0.65&   0.80 &   0.71&   0.75 \\
$\Lambda$&$\to$&$p$
&$-$0.98 &$-$1.08 &$-$0.96&$-$0.89 &$-$0.87&$-$0.89 \\
$\Xi^-$&$\to$&$\Lambda$
&   0.37 &   0.38 &   0.35&   0.17 &   0.40&   0.16 \\
$\Sigma^-$&$\to$&$n$
&   0.27 &   0.32 &   0.27&   0.50 &   0.40&   0.49 \\
\hline
\multicolumn{3}{c}{$\chi^2/N$}
& 14.7  & 1026 & 8.5 & 13.2  & 3.5  & 6.5 \\
\end{tabular}
\caption{The axial-vector coupling constants for the measured
semileptonic decays in different models: octet baryons with
pseudoscalar mesons $(oct,ps)$, octet and decuplet baryons with
pseudoscalar mesons $(ps)$ and with vector mesons in addition
$(all)$.  The axial coupling constants for the $SU(3)$ case are:
$F=0.53$, $D=1.15$ for $(oct,ps)$ and $F=0.48$, $D=0.91$ for $ps$
and $F=0.53$, $D=1.15$ for $all$.}
\end{table}

\begin{table}
\begin{tabular}{rccccc}
                & $S_{EJ}^p$ & $S_{EJ}^n$ & $S_B$
& $\Delta q_0$ & $\Delta s$ \\
\hline
$SU(6)\>\>tree$ & 0.278      &   0.       & 0.278
&  1.          & 0.      \\
   $oct,ps$     & 0.212      &   0.004    & 0.208
&  0.779       & 0.004   \\
   $ps$         & 0.233      &$-$0.010    & 0.243
&  0.804       & 0.002   \\
   $all$        & 0.220$^{*)}$ &   0.011    & 0.209$^{**)}$
&  0.846       & 0.017   \\
\hline
$SU(3)\>\>tree$ & 0.173      &$-$0.037    & 0.210
&  0.489       & 0.      \\
  $oct,ps$      & 0.154      &$-$0.056    & 0.210
&  0.356       & 0.003   \\
  $ps$          & 0.169      &$-$0.041    & 0.210
&  0.461       & 0.001   \\
  $all$         & 0.179$^{*)}$ &$-$0.031    & 0.210$^{**)}$
&  0.541       & 0.018   \\
\hline
\hline
 $EMC$ \cite{A89} & 0.126  & -- & -- & 0.120 &  $-$0.19  \\
$[Q^2=10.7GeV^2]$
 & $\pm 0.01 \pm 0.015$ & & &
 $\pm 0.094 \pm 0.138$ & $\pm 0.032 \pm 0.046$ \\
\hline
 $E142$ \cite{AAB93} & -- & $-$0.022  & 0.146 & 0.57  & $-$0.01   \\
$[Q^2=2GeV^2]$
 & & $\pm 0.011$ & $\pm 0.021$ (+EMC) & $\pm 0.11$ & $\pm 0.06$ \\
\hline
  $SMC$ \cite{AAA94} & 0.136  & -- & -- & 0.22  & $-$0.12  \\
$[Q^2=5GeV^2]$
 & $\pm 0.011 \pm 0.011$ & & &
 $\pm 0.10 \pm 0.10$ & $\pm 0.04 \pm 0.04$ \\
\hline
$all$ \cite{AAA94}   & 0.142 & $-$0.069 & 0.204  & 0.27 & $-$0.10  \\
$[Q^2=5GeV^2]$
 & $\pm 0.008 \pm 0.011$ & $\pm 0.025$ &
 $\pm 0.029$ & $\pm 0.08 \pm 0.10$ & $\pm 0.03 \pm 0.04$ \\
\end{tabular}
\caption{The Ellis-Jaffe Sum Rule for proton and neutron, the Bjorken
Sum Rule and the axial flavour singlet constant obtained with
inclusion of different Fock states (see text). In the lowest panel we
have collected recent experimental data of EMC, E142 and SMC
collaborations. The theoretical values of $S_{EJ}$ and $S_{B}$ are
not corrected for pQCD (!).\hfil\break $^{*)}$ For $Q^2$ = 5 GeV$^2$,
$S_{EJ}^{p}$ corrected for pQCD effects according to
\protect\cite{K80} is 0.21 ($SU(6)$) and 0.17 ($SU(3)$).\hfil\break
$^{**)}$ The pQCD corrections up to $\alpha_{s}^3$ give
correspondingly $S_{B}$ = 0.19 in both $SU(6)$ and $SU(3)$ model.}
\end{table}

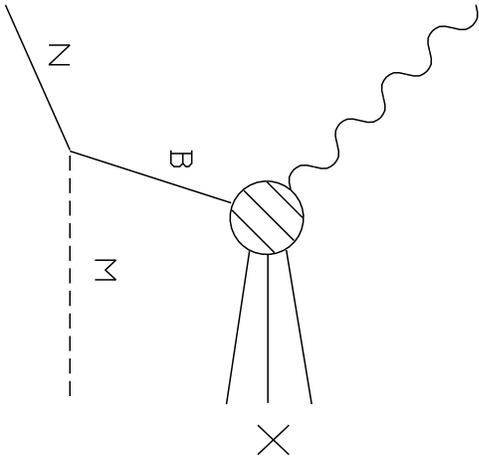
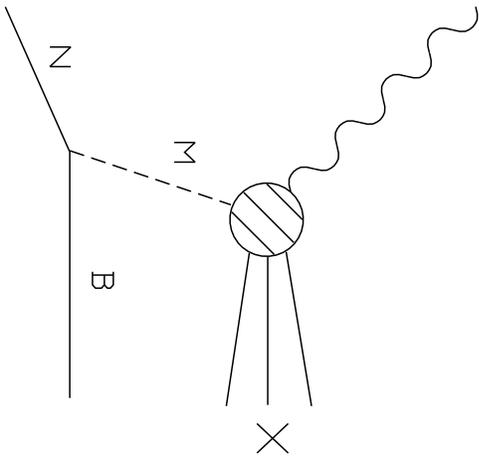
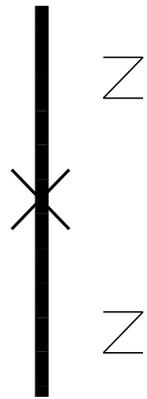
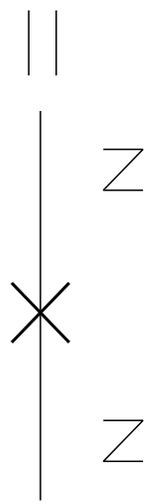
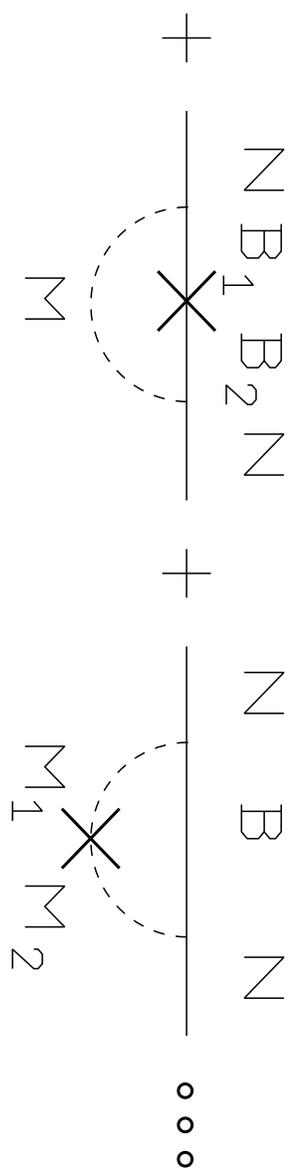
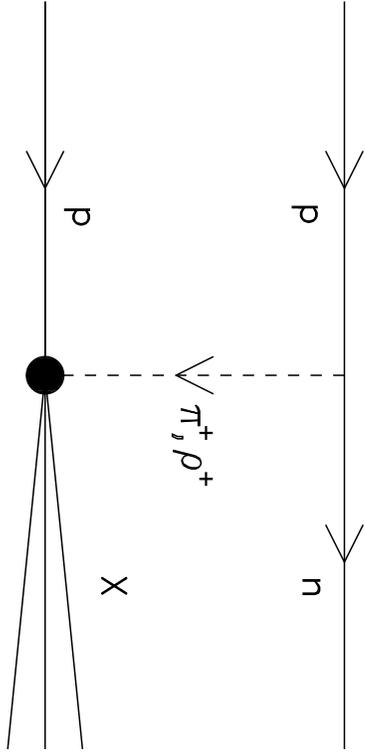
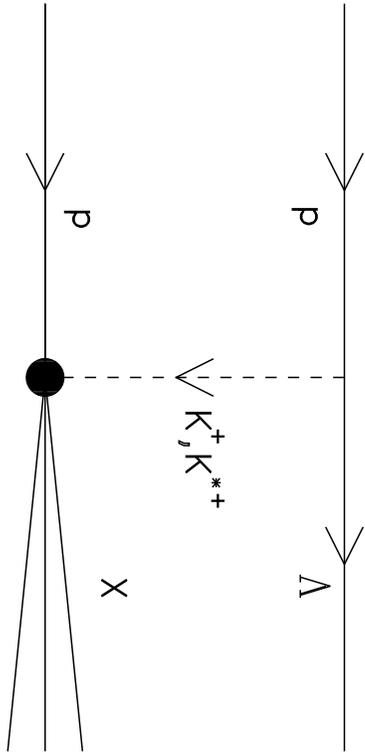
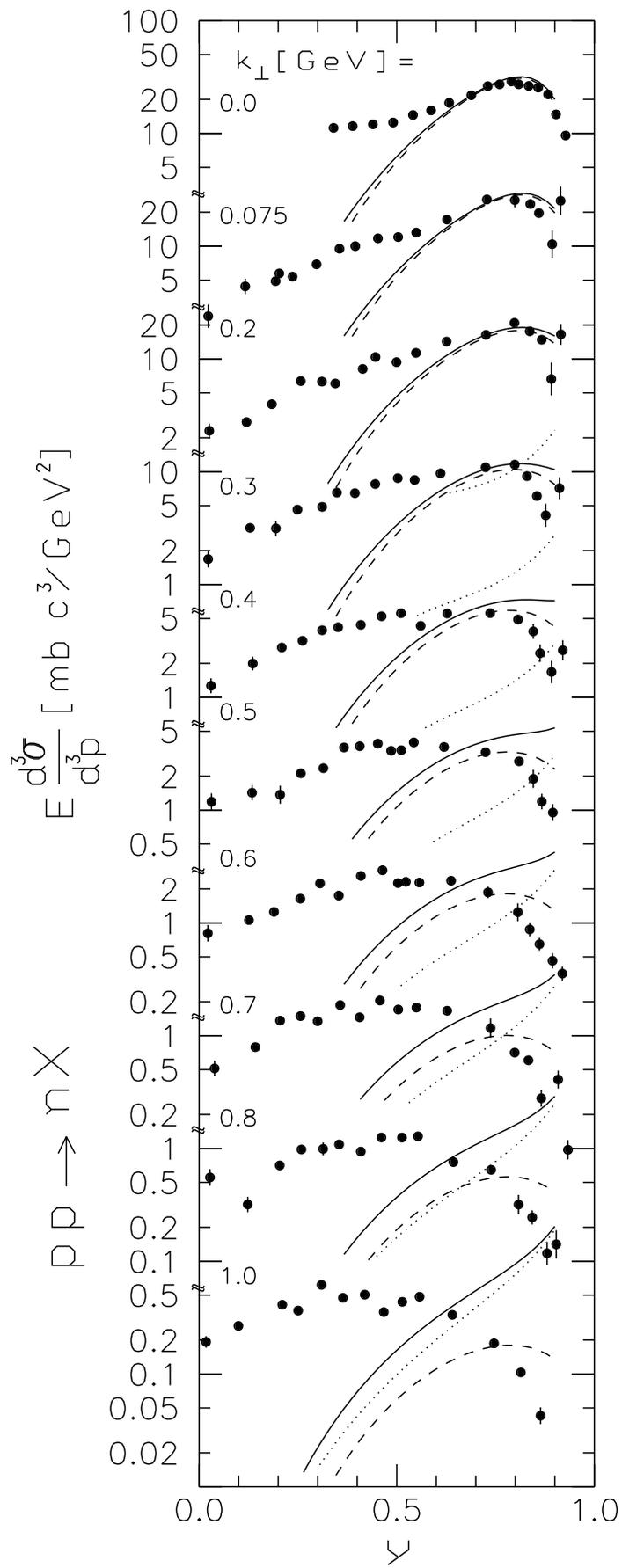
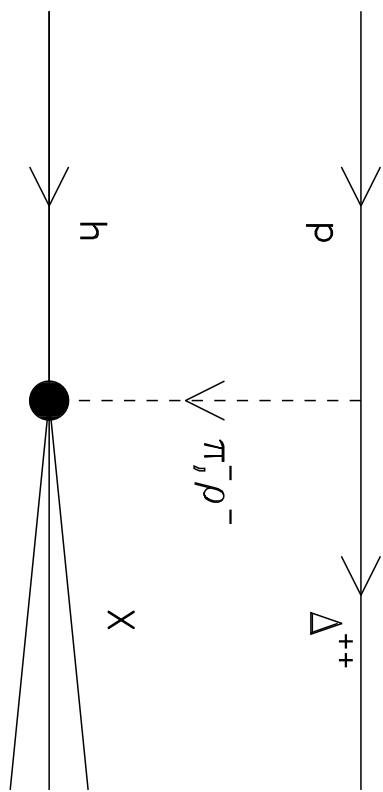
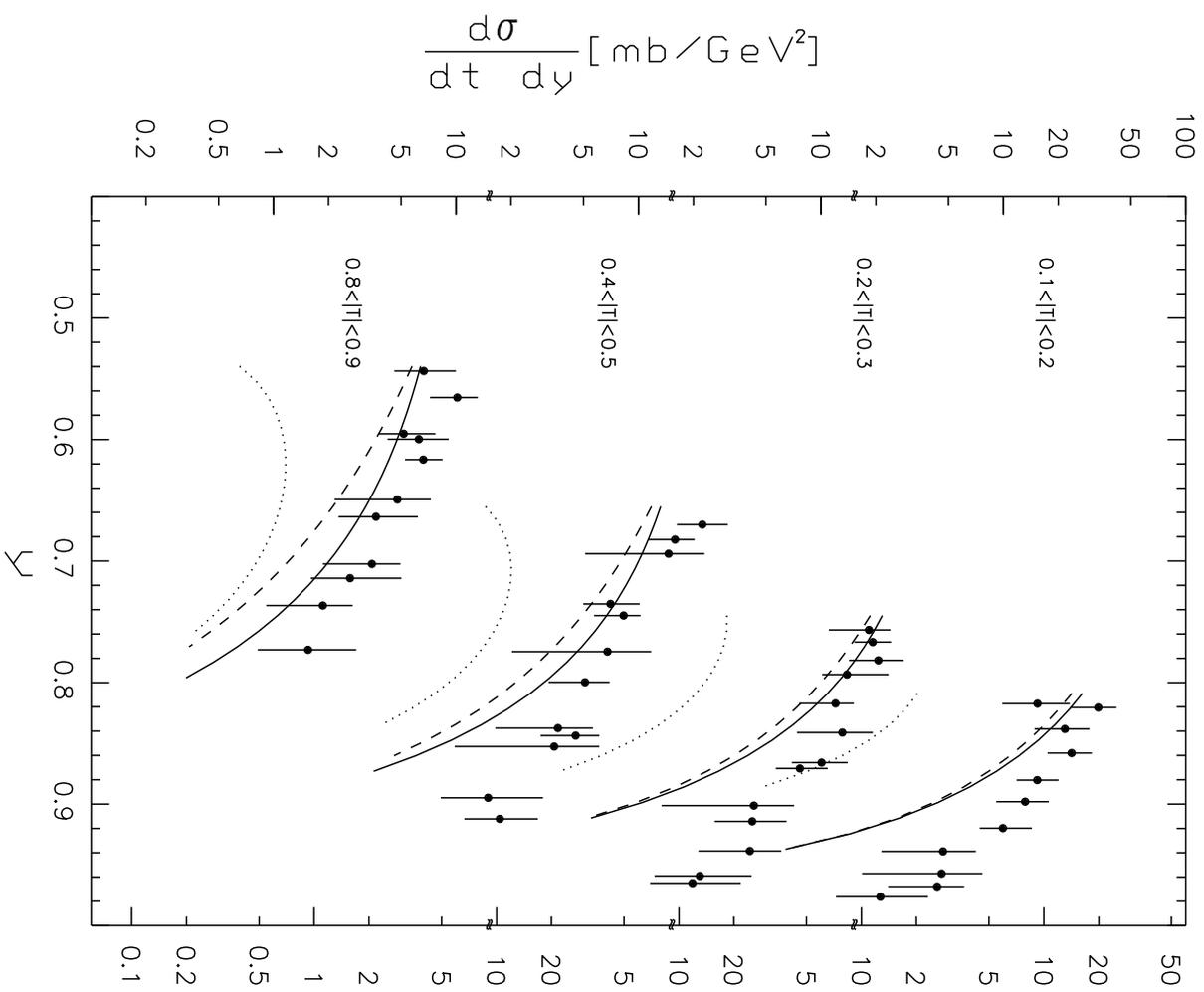
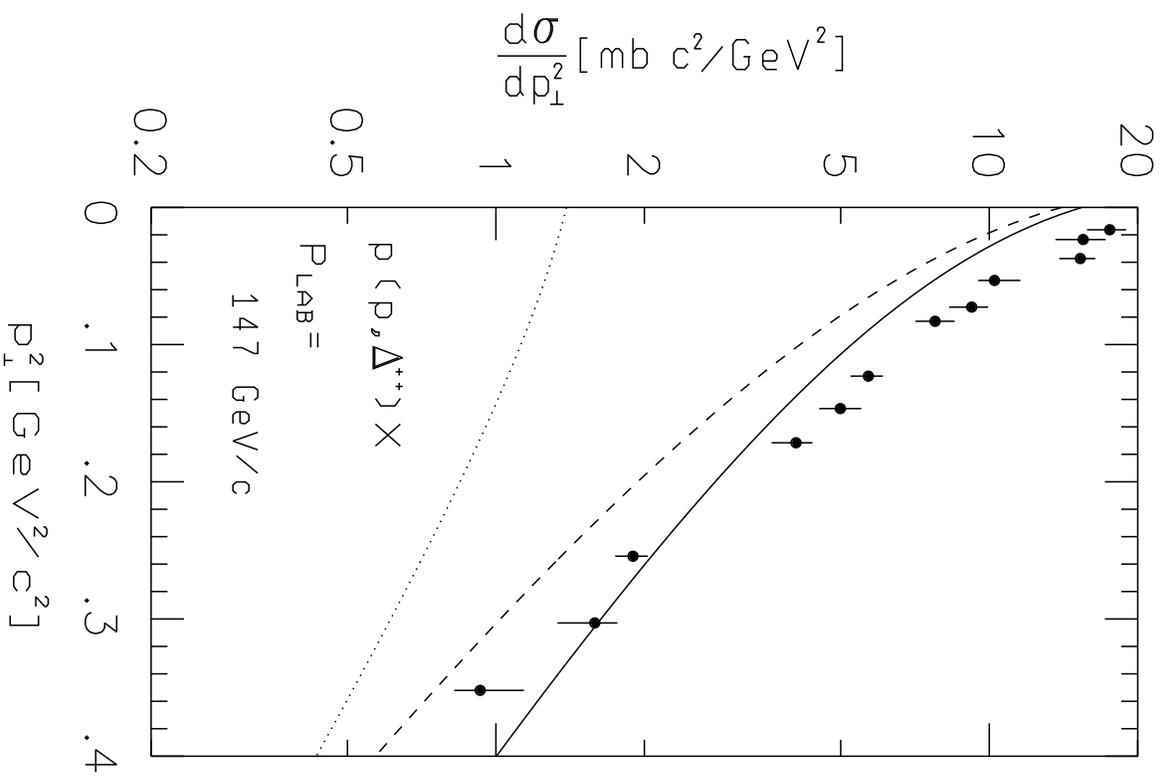
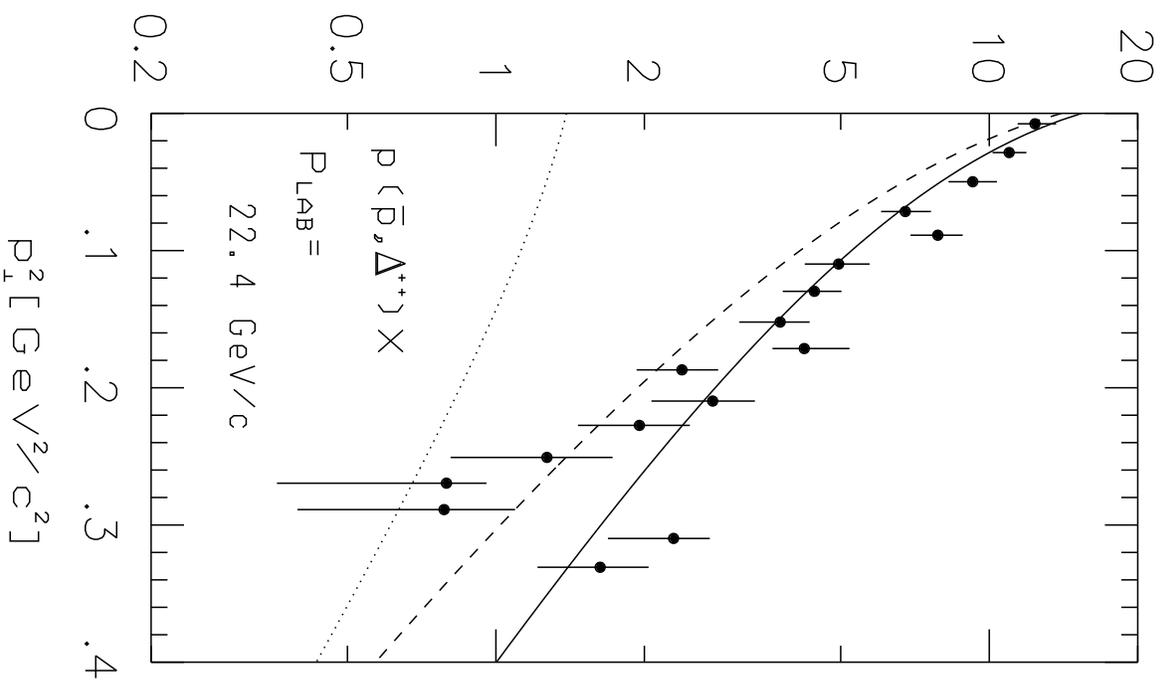
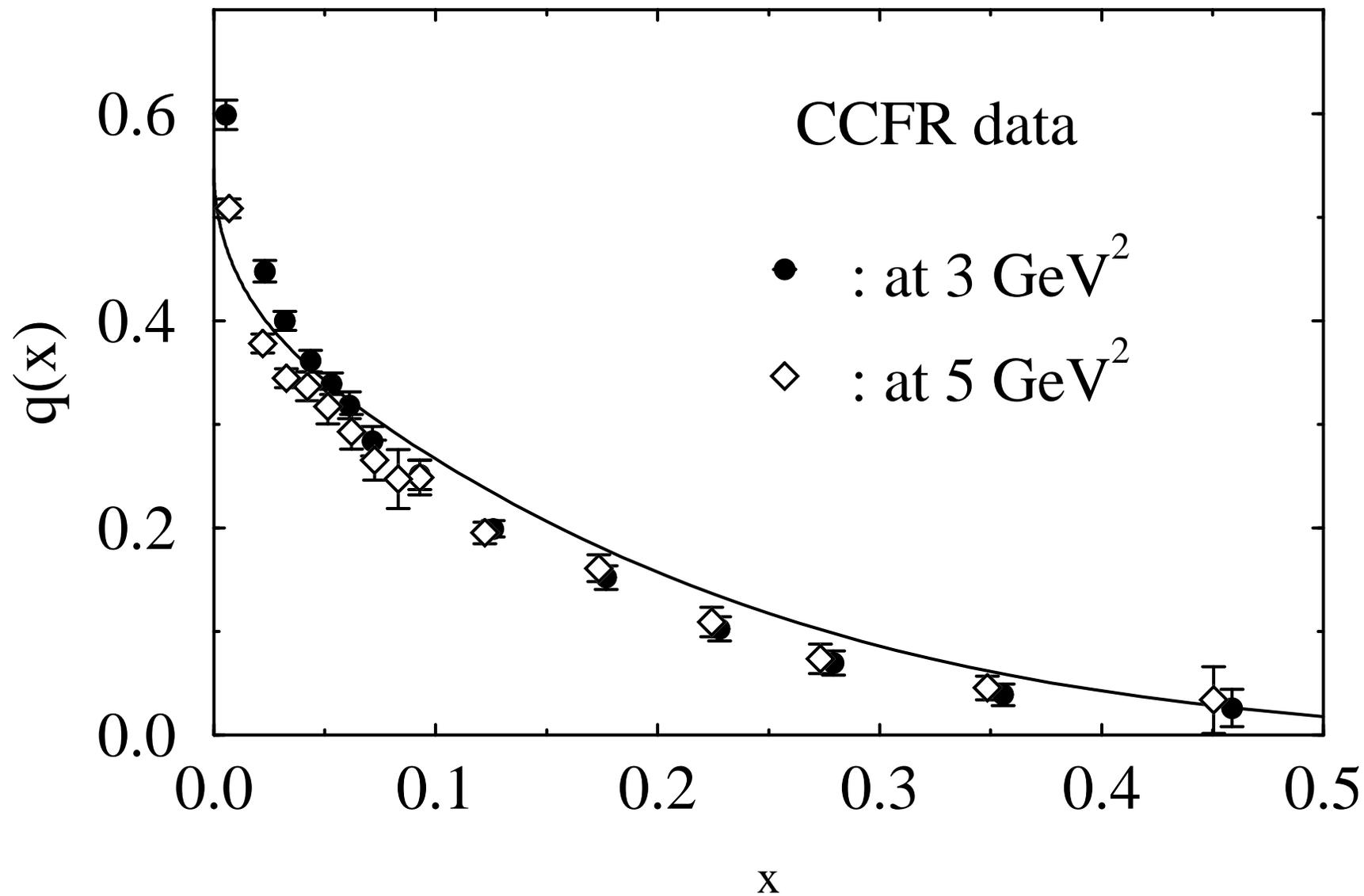
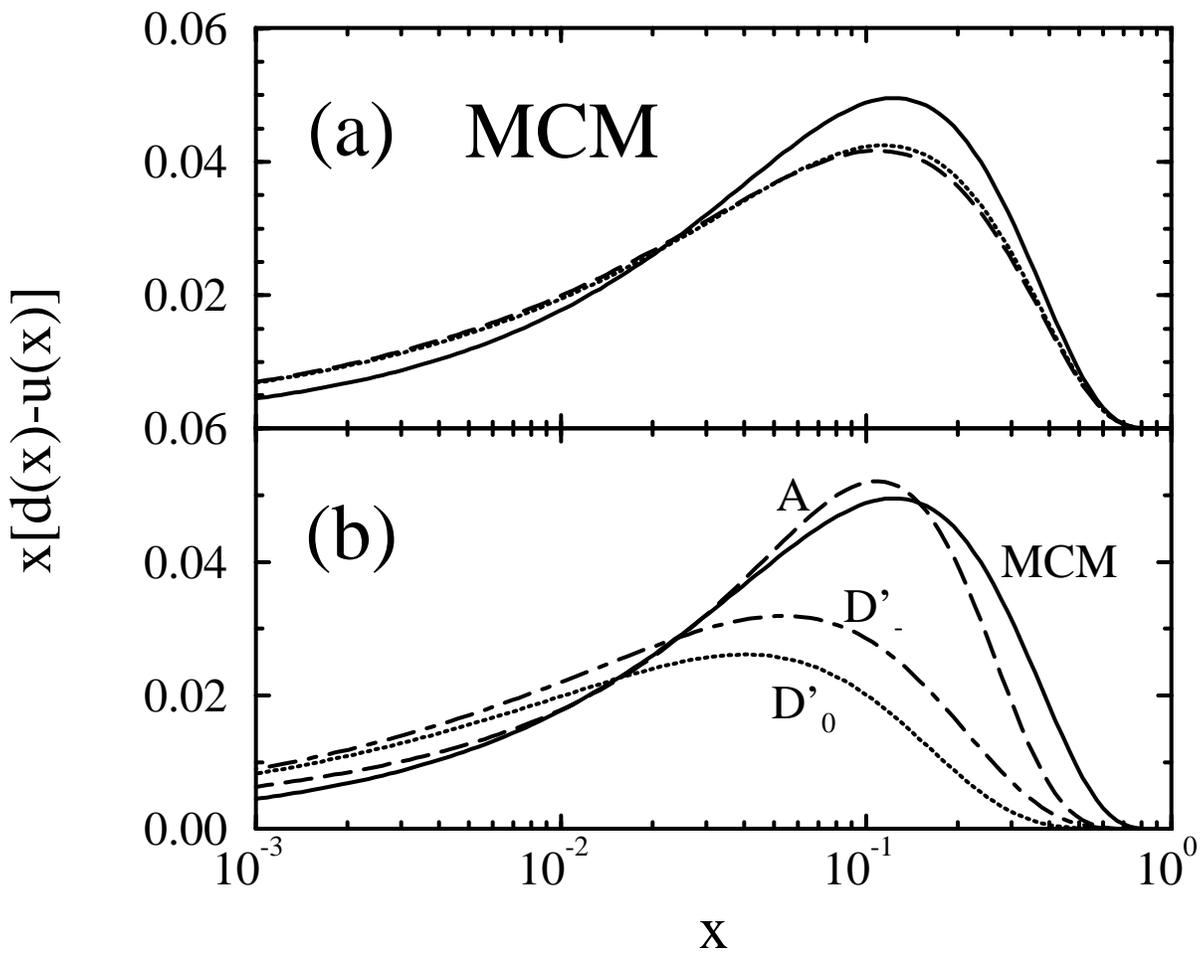
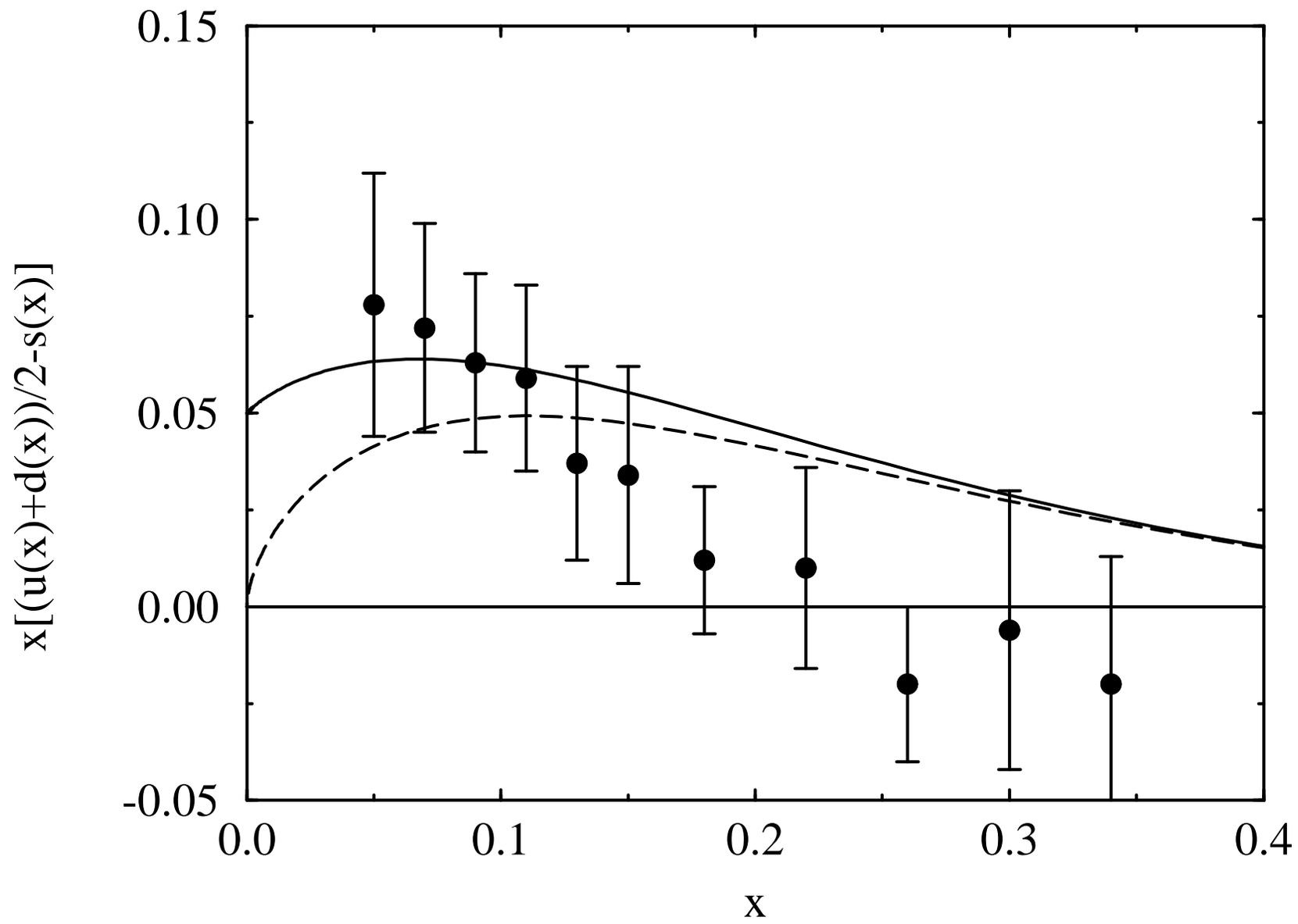
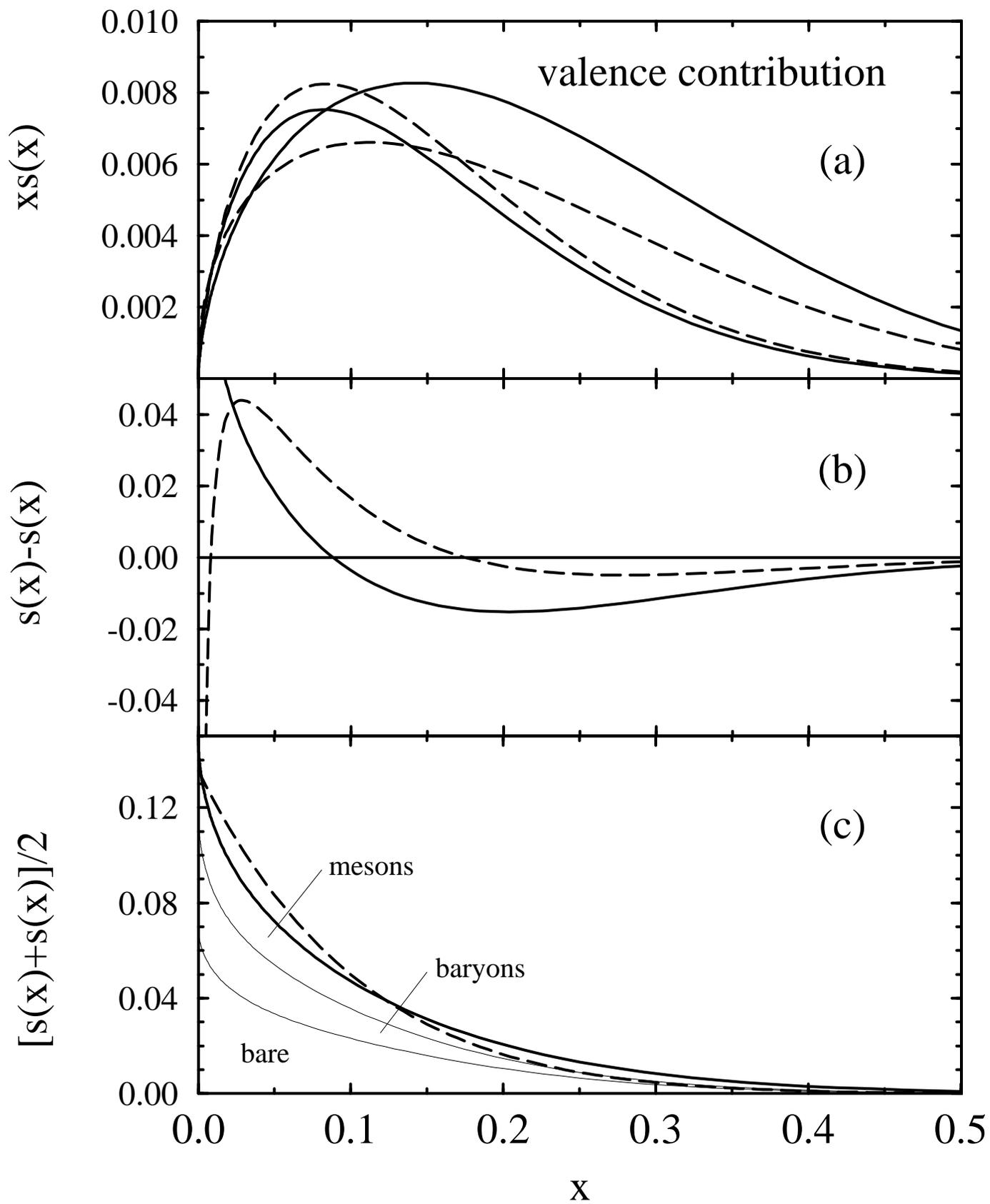
\end{document}